# All-Optical Control of Interfacial Polarization in $MoS_2/WSe_2$ Heterobilayers

Muhammad Sufyan Ramzan,[1*] Giancarlo Soavi,[2,3] and Caterina Cocchi[1,3*]

[1] *Institut für Festkörpertheorie und Optik, Friedrich-Schiller-Universität Jena, 07743 Jena, Germany*

[2] *Institut für Festkörperphysik,[2] Friedrich-Schiller-Universität Jena, 07743 Jena, Germany*

[3] *Abbe Center of Photonics, Friedrich-Schiller-Universität Jena, 07745 Jena, Germany*

*Email:* muhammad.ramzan@uni-jena.de, caterina.cocchi@uni-jena.de

**Abstract:**

All-optical tuning of van der Waals (vdW) heterostructures with coherent radiation offers a powerful strategy for manipulating interfacial electronic properties on femtosecond timescales. Using real-time-time-dependent density functional theory, we predict a transient out-of-plane polarization in $MoS_2/WSe_2$ heterobilayers driven by intense ultrafast pulses. In the weak-field regime, the system preserves its intrinsic type-II band alignment, whereas at intermediate intensities, the field triggers a nonlinear enhancement of interlayer charge transfer. High-harmonic generation spectra undergo a crossover from the perturbative to the strong-field regime accompanied by the pronounced enhancement of photoinduced interfacial dipole. At the highest intensities, interlayer polarization enhancement is sustained by massive field-driven spatial delocalization of electronic density into the vdW gap, promoted by localized charge saturation due to Pauli blocking. Lattice strain modifies the resonance conditions while preserving this strong-field polarization enhancement, establishing a microscopic mechanism for all-optical manipulation of interfacial dipoles at the ultrafast scale.

**Main text**

The control of electronic phases in van der Waals (vdW) heterostructures via light-matter interactions has emerged as a cornerstone for the development of quantum materials [1–3]. Typically, tuning the optical properties of these systems is achieved via static approaches, such as chemical doping [4,5] or the formation of polariton states within optical cavities [6–9]. A compelling alternative for achieving light-induced functional changes is offered by nonlinear optics: while weak excitations maintain the system within linear response, where higher-order harmonic signals are negligible, intense pulses drive the system into the non-perturbative, strong-field regime [10–13]. In this limit, the interaction triggers the formation of hybrid light-matter states [14], inducing non-equilibrium phase transitions [15–17]. Such all-optical control bypasses the need for complex nanophotonic architectures and enables ultrafast manipulation of interfacial dipoles and non-perturbative electronic transport on femtosecond timescales [7,18,19].

Transition metal dichalcogenide (TMD) heterobilayers are ideal quantum-material platforms for these investigations. Beyond their chemical stability and structural flexibility [20–22], they offer a tunable excitation landscape across the entire visible region [23–27]. The type-II band alignment, achieved with specific material combinations [28–30], promotes efficient interlayer charge separation [29,31]. Recently, photoinduced charge transfer in polar TMD homobilayers has been predicted to drive ultrafast out-of-plane polarization changes purely through electronic mechanisms rather than lattice distortions [32]. Furthermore, the inherent lattice mismatch among TMD monolayers introduces strain as an additional degree of freedom in these heterostructures [33,30,34]. While spatial inhomogeneities typically fade out in experiments, atomistic simulations have revealed that local strain domains act as high-precision tuning knobs to engineer the properties of the materials and their ability to interact with light [35,36].

In this work, we demonstrate the emergence of laser-induced interfacial polarization in $MoS_2/WSe_2$ heterobilayers using real-time time-dependent density-functional theory (RT-TDDFT) simulations. This first-principles approach allows us to track the femtosecond (fs) evolution of the electronic density simultaneously with the field propagation, capturing transient out-of-plane polarization response persisting post-pulse. By investigating the dynamics in the absence of non-electronic dissipation channels, we isolate pure electronic coherence. In the nonlinear regime, we identify an intermediate intensity enhancing interlayer charge transfer by a factor of approximately 2.5. Under even stronger field intensities, the response of the system becomes non-perturbative, leading to a transient state characterized by a significantly enhanced out-of-plane dipole. Furthermore, the demonstrated robustness of this effect with respect to strain offers a controllable knob for fine-tuning the resonance condition without compromising interfacial dipole enhancement.

We begin by establishing the linear optical response of the considered $MoS_2/WSe_2$ heterobilayer (Figure 1a) as the baseline for our investigation. The characteristics of the linear absorption spectra remain qualitatively invariant across different strain distributions, namely averaged between the two monolayers or entirely allocated on either of them (Figure 1b). Specifically, the primary intense resonance at ~2.75 eV dominates all spectra. Microscopically, this peak arises predominantly from transitions involving interlayer hybridized states rather than simple localized transitions (Figure S1). Its oscillator strength is maximized under an equally distributed average strain, while remaining only marginally weaker when strain is localized to a single monolayer. The other spectral features are more sensitive to structural variations, directly mirroring corresponding changes in the band structure (Figure S1)[36].

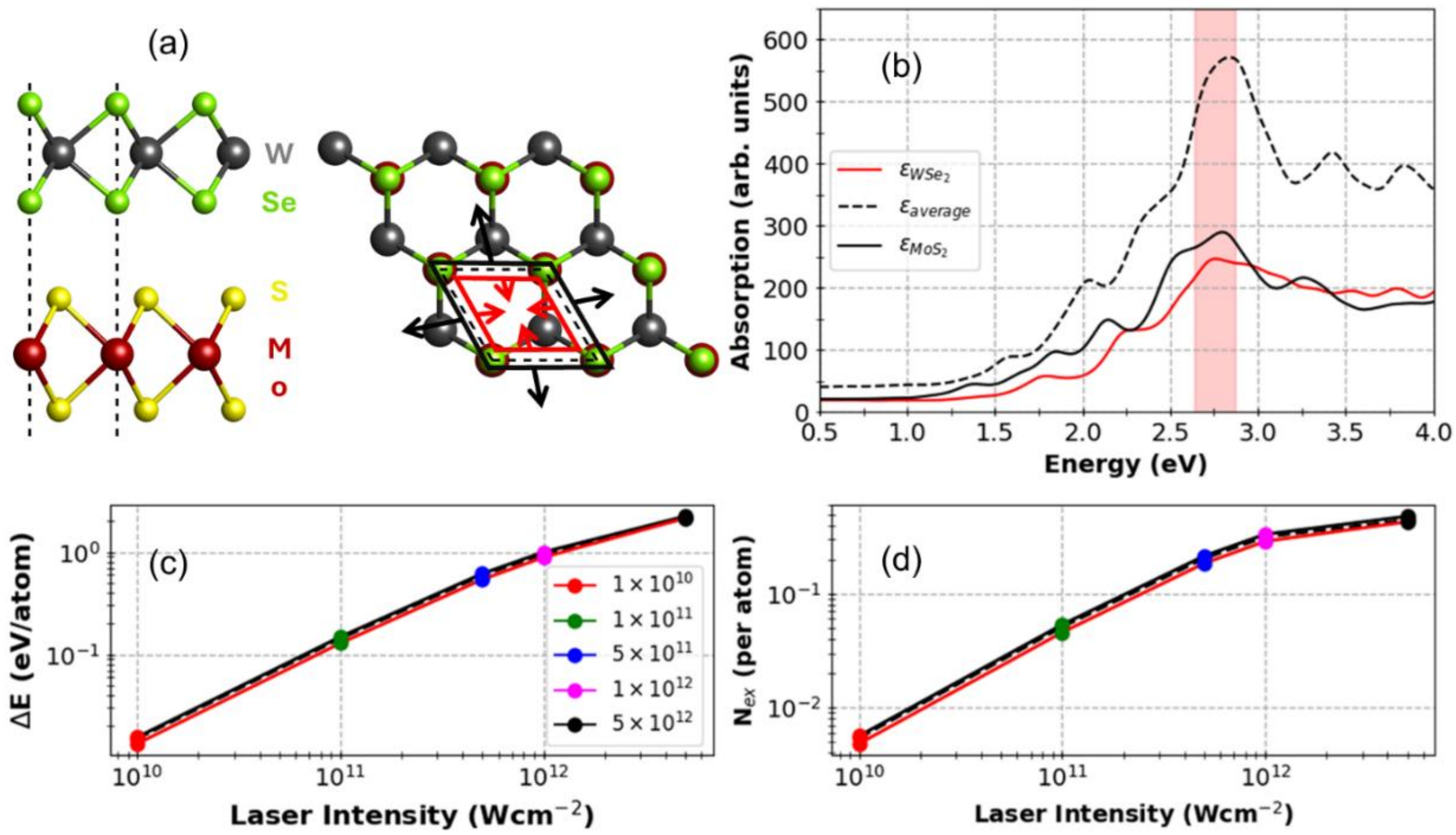

**Figure 1.** (a) Side and top view of the crystal structure of the $MoS_2/WSe_2$ heterobilayer. Red solid, black solid and black dashed rectangles mark the unit cells with in-plane lattice strain applied entirely to monolayer (ML) $WSe_2$, to ML $MoS_2$ or equally distributed between both layers, respectively. (b) Linear absorption spectrum for the three considered heterostructures with the solid black and red curves representing the systems where the strain is only on $MoS_2$ and $WSe_2$, respectively. The shaded area highlights the spectral region in resonance with the applied pulses. (c) Absorbed energy per atom and (d) number of excited electrons per atom at the end of the simulation (t = 50 fs) under increasing laser intensities.

We find a spectral shift of the absorption onset of roughly 0.4 eV between the $MoS_2$-strained (1.4 eV) and $WSe_2$-strained (1.8 eV) configurations. This tunability is critical, demonstrating that strain can be utilized to align the (resonant) response of the material to specific laser frequencies without substantially altering its overall spectral profile [37,38]. Importantly, compared to traditional chemical strategies such as alloying [39], which involve complex material synthesis and potential structural defects, strain engineering offers a non-invasive, reversible, and highly localized alternative for tuning the optoelectronic landscape of vdW heterostructures [36].

For a consistent analysis of laser-induced dynamics, we set the pulse carrier frequency in resonance with the maximum at 2.75 eV (Figure 1b). By varying the field intensity from $1\times10^{10}$ Wcm$^{-2}$ to $5\times10^{12}$ Wcm$^{-2}$, we monitor the total energy uptake (ΔE) and the density of excited carriers ($N_{ex}$) to identify three regimes of excitation that are nearly independent of the strain condition (Figure 1c-d and Figure S2). While saturable absorption is known to appear at intensities below $10^{10}$ – $10^{11}$ Wcm$^{-2}$ [40], in this regime, the response remains dominated by linear effects, as testified by the trends of both ΔE and $N_{ex}$ (Figure S2). In the intermediate intensity regime (I = $10^{11}$

– $10^{12}$ $Wcm^{-2}$), the energy uptake scales sublinearly ($\sim I^{0.8}$), indicating the transition from linear response to partial state filling. Above $10^{12}$ $Wcm^{-2}$, the energy uptake rises dramatically, with almost 0.5 electrons promoted to the excited state. In this non-perturbative regime, both ΔE and $N_{ex}$ exhibit strongly sublinear scaling ($\sim I^{0.5}$ or lower), signaling the onset of saturable absorption. It is worth noting that while the absolute intensities marking the onset of nonlinear saturation are sensitive to the details of the RT-TDDFT simulations (e.g., real-space grids, simulation box, etc.) and the inclusion of explicit dissipation channels, the qualitative trends are robust and physically representative of the response of the $MoS_2$/$WSe_2$ heterobilayer to high fields.

To gain additional insight into the response of the systems in these different excitation regimes, we examine the macroscopic current density in the frequency domain (Figure 2; time-domain signals in Figure S3). Given the qualitatively identical behavior of the heterostructures under different strain conditions, we focus on the heterobilayer with averaged strain, providing the corresponding results for the other two systems in the Supporting Information (Figures S4 and S5). Under the weakest intensity of $1\times10^{10}$ $Wcm^{-2}$, the spectral response is dominated by the fundamental frequency $\omega_0$ = 2.75 eV (Figure 2a). While nonlinear optical processes are inherently present at all field strengths, the intensity of higher-order harmonics in this weak-field regime are not detectable in our simulations. As the intensity increases to the $10^{11}$ - $10^{12}$ $Wcm^{-2}$ range, the second and third harmonics emerge prominently (Figure 2b-d).

We note that the broken inversion symmetry of the $MoS_2$/$WSe_2$ heterostructure activates the even-harmonics [41,42]. While discrete peaks related to individual high harmonics remain visible (Figure 2e), the emergence of a broad spectral background marks the transition from isolated harmonic generation to a regime of strong electronic renormalization, where the external field significantly dresses the quasiparticle states, facilitating the observed phase transition. To quantify this filling, we define the background-to-peak ratio as

$$\eta_{\text{bg}} = I_{\text{midpoint}} / \max(I_{\text{peak}}),$$

where $I_{\text{midpoint}}$ is the median spectral intensity between adjust harmonics and $I_{\text{peak}}$ is the maximum intensity of the corresponding harmonics. For low-order intervals ($1\omega_0 - 2\omega_0$ and $2\omega_0 - 3\omega_0$), $\eta_{\text{bg}}$ increases by over two orders of magnitude, from $\sim 3 \times 10^{-4}$ to $> 10^{-1}$ (Figure 2f), reflecting non-perturbative continuum filling and sub-cycle electronic redistribution into the vdW gap. For higher-order intervals ($3\omega_0 - 6\omega_0$), $\eta_{\text{bg}}$ remains bounded below 0.2, confirming that discrete harmonic peaks remain sharp and coherent across the entire intensity range. In the strong field regime ($1\times10^{12}$ - $5\times10^{12}$ $Wcm^{-2}$), the HHG spectrum undergoes a marked redistribution of spectral weight, with discrete low-order harmonics progressively merging into a continuous background.

Equipped with the knowledge on the (nonlinear) response of the system to different field intensities, we move on to the analysis of laser-driven charge-transfer across the interface. As visualized in Figure 3, all considered systems exhibit nonzero charge transfer in the static regime (t =0), with $MoS_2$ accepting electrons from $WSe_2$, in accordance with the electronic structure (Figure S1). In the weak-field regime (I = $1\times10^{10}$ $Wcm^{-2}$), photoinduced electron population migrates from $WSe_2$ to $MoS_2$, consistent with the type-II level alignment (Figure S1) [30]. The resulting change in charge population reported in Figure 3a is of the order of 0.01 electrons. Notably, residual oscillations of the charge imbalance following the pulse are an order of magnitude smaller than the net induced transfer. In the absence of dissipation channels (fixed ions, no energy loss), this charge transfer remains largely constant for the entire propagation duration. Nuclear motion is not expected to fundamentally alter this electronically established mechanism, as discussed in previous work for TMD-based hybrid interfaces using ensemble average of nuclear trajectories and quantum nuclear effects [43].

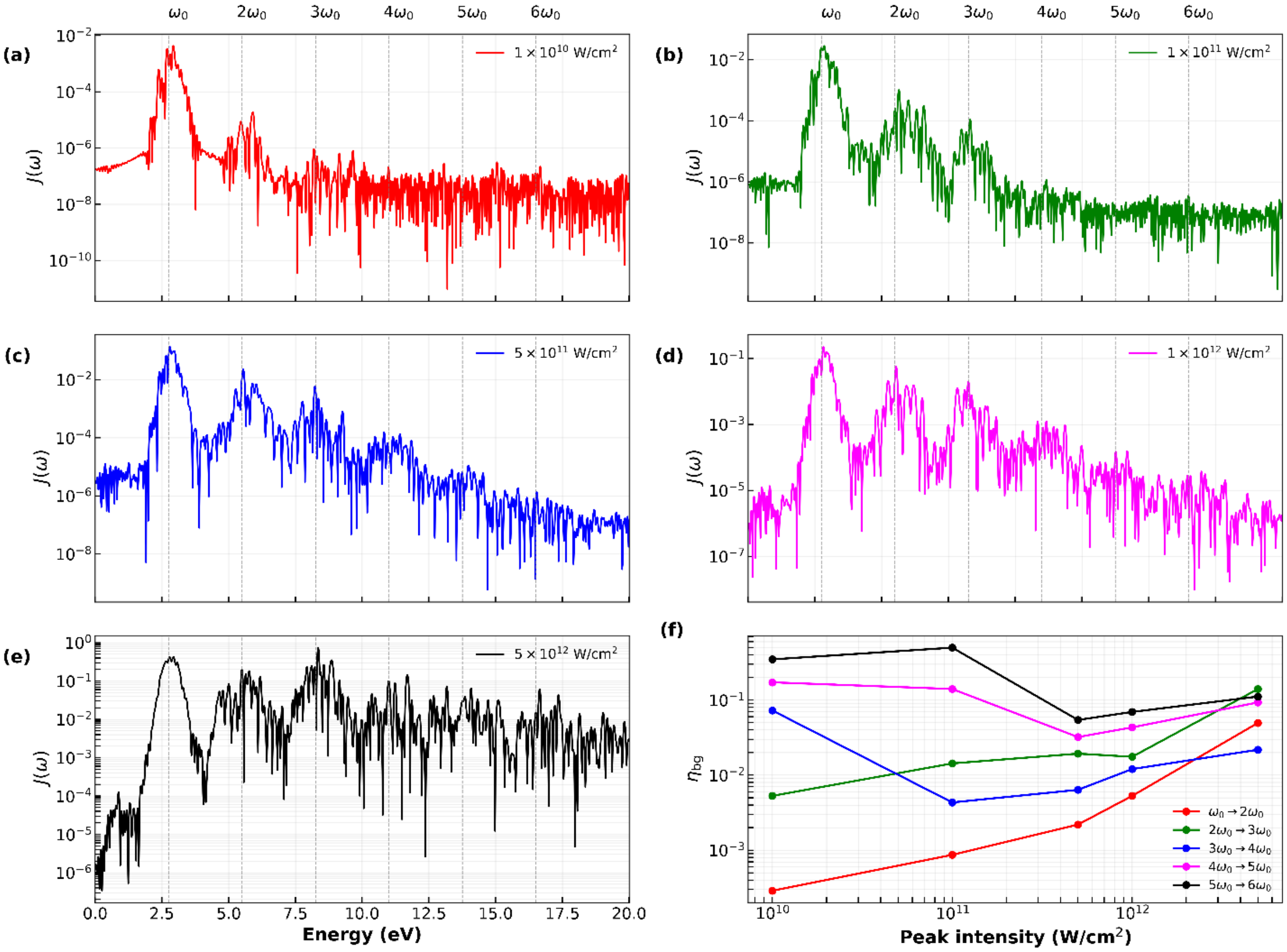

**Figure 2**. High-harmonic spectrum of the $MoS_2/WSe_2$ heterobilayer with equally distributed strain under (a) $1\times10^{10}$ $Wcm^{-2}$, (b) $1\times10^{11}$ $Wcm^{-2}$, (c) $5\times10^{11}$ $Wcm^{-2}$, (d) $1\times10^{12}$ $Wcm^{-2}$, and (e) $5\times10^{12}$ $Wcm^{-2}$. The fundamental carrier frequency ($\omega_0$) and its overtones are indicated by dashed bars. (f) Midpoint background-to-peak ratio, $\eta_{bg}$, as a function of driving laser intensity for adjacent harmonic intervals up to the 6th order.

Moving to the intermediate intensity regime ($10^{11}$ - $10^{12}$ $Wcm^{-2}$), we notice an increased net charge transfer after irradiation, with a mean value of about 0.025 electrons (Figure 3b-d). While nonlinear, this regime remains qualitatively consistent with the perturbative response, characterized by reversible charge fluctuations that plateau post-pulse due to the absence of explicit dissipation channels. Notably, even at a laser intensity of $10^{12}$ $Wcm^{-2}$, a pronounced decrease in charge transfer is visible (Figure 3d), indicating the onset of Pauli blocking. The situation changes substantially at the highest intensity ($5\times10^{12}$ $Wcm^{-2}$), where the system enters a non-equilibrium regime governed by Pauli blocking and state saturation (Figure 3e) already predicted in a hybrid TMD-based interface [44]. During pulse duration, the field rapidly promotes a dense carrier concentration that fills the localized conduction states of the constituents.

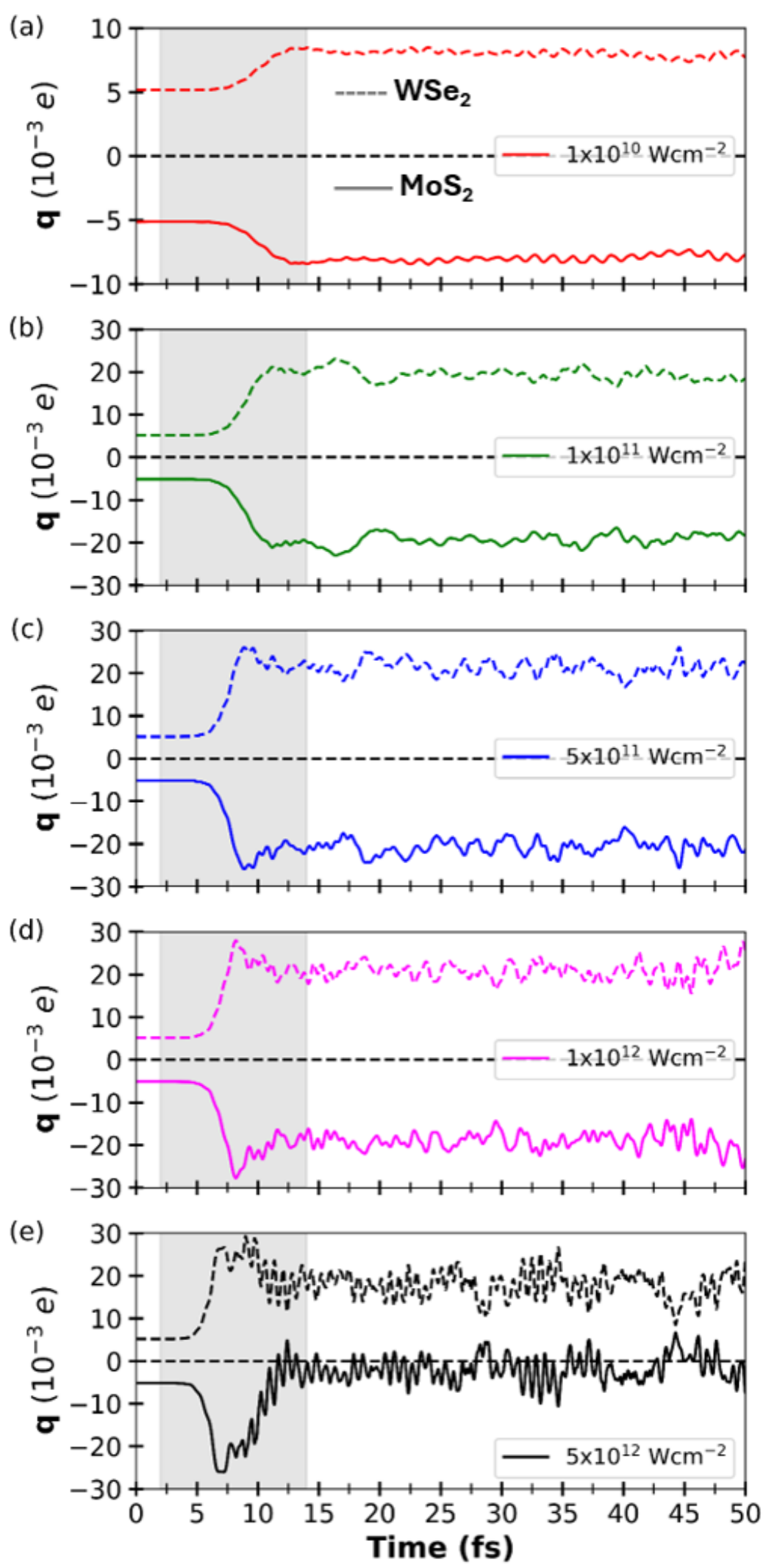

**Figure 3.** Time evolution of the layer-projected atomic charge (q) on each layer of the $MoS_2/WSe_2$ heterostructure with equally distributed strain compared to the isolated monolayer, under the laser intensities of (a) $1\times10^{10}$ $Wcm^{-2}$, (b) $1\times10^{11}$ $Wcm^{-2}$, (c) $5\times10^{11}$ $Wcm^{-2}$, (d) $1\times10^{12}$ $Wcm^{-2}$, and (e) $5\times10^{12}$ $Wcm^{-2}$. Negative (positive) values of q indicate charge accumulation (depletion) shown with solid (dashed) line. The gray shaded area marks the time window of laser irradiation.

Once these available channels are saturated, further net accumulation of localized charge is suppressed, forcing the layer-projected Hirshfeld charge of $MoS_2$ to oscillate symmetrically around zero net transfer post-irradiation. Crucially, the absence of atomic charge accumulation does not imply a lack of electronic reconfiguration. Instead, the intense field drives the filled states to undergo a massive spatial expansion, delocalizing the electronic density directly into the interstitial vdW gap and vacuum regions. This charge transfer mechanism is independent of the in-plane strain (see Figure S6). The duration and reversibility of this transient state in an experimental scenario is subject to the interplay of the quantum system with intrinsic and extrinsic dissipation channels [45–47]. However, the results of our simulations demonstrate that the interaction with a sufficiently intense field promotes the system to a non-equilibrium configuration with amplified out-of-plane polarization, effectively reshaping the electronic landscape without the need for an external cavity or resonating electromagnetic environments [48–50].

Finally, we quantify this effect by analyzing the time evolution of static out-of-plane electronic dipole moment, calculated by integrating the time-dependent electronic charge density $\rho(\mathbf{r},t)$ over the simulation cell volume:

$$p_z(t) \; = \; \int_\Omega \rho(\boldsymbol{r},t) z d\boldsymbol{r}^{\mathbf{3}},$$

where *z* represents the normal coordinate to the heterobilayer and Ω is the volume of simulation cell. Since the supercell remains charge-neutral even under strong-field conditions, the total integrated charge is zero ($\int_\Omega \rho(\boldsymbol{r},t) d\boldsymbol{r}^{\mathbf{3}} = 0$), making the dipole invariant to the choice of origin. It is worth noting that the while the Hirshfeld charge profile remains almost invariant to the vacuum layer size even at the strongest field intensity, the sensitivity in the polarization cannot be eliminated, due to the diffuse electronic variation by about 20% in the post-pulse dipole (Figure S7). However, the overall increase in out-of-plane dipole and partial charge-transfer saturation remain robust across all tested vacuum sizes.

In the steady state, the heterobilayer exhibits a dipole of 0.51 D, emerging from the interlayer charge distribution (Figure 4). Under the weakest irradiation ($10^{10}$ Wcm$^{-2}$), the interfacial dipole increases slightly to approximately 0.63 D. In the intermediate regime ($10^{11}$ – $10^{12}$ Wcm$^{-2}$), we observe a marked non-linearity: while the carrier density $N_{ex}$ scales quasi-linearly with intensity, the induced dipole moment increases by a factor of ~2.5, suggesting this quantity is enhanced by the field-induced modification of the interlayer electronic potential. Under the strongest laser pulse (I = 5×$10^{12}$ Wcm$^{-2}$), the out-of-plane dipole increases by nearly a factor of 6. This behavior reconciles the apparent paradox of a vanishing Hirshfeld layer charge (Figure 3e) taking place alongside a maximized dipole moment. According to Eq. (1), the total dipole moment $p_z(t)$ accounts for the full spatial distribution $\rho(\boldsymbol{z},t)$. Because Pauli blocking prevents electrons from nesting tightly within the localized atomic spheres of the layers, the density is driven out into the interstitial gap and vacuum boundaries.

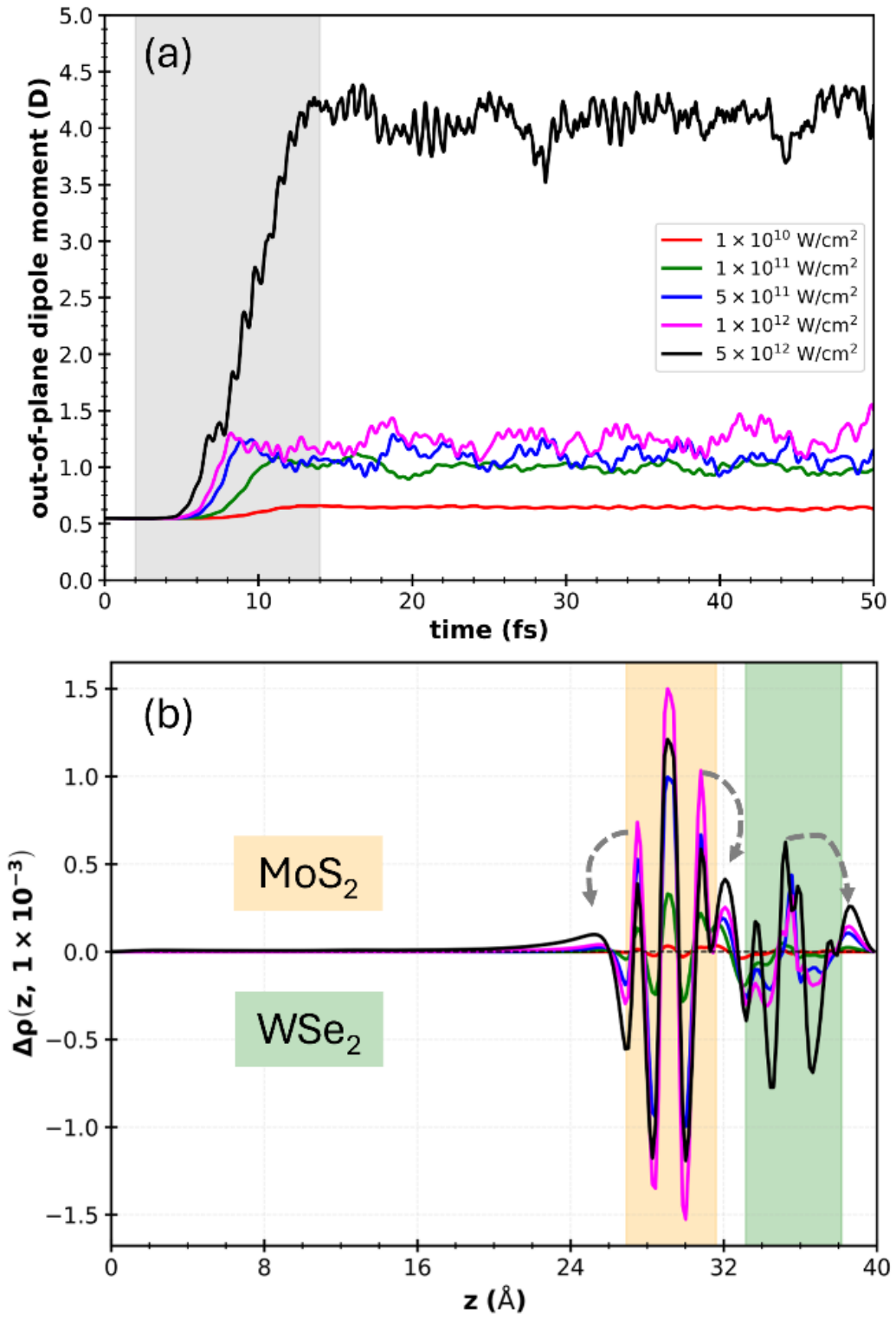

**Figure 4.** (a) Time evolution of the out-of-plane dipole moment and (b) planar-averaged induced charge density different, $\Delta\rho(z)$, along z vector of the $MoS_2/WSe_2$ heterobilayer with equally distributed strain under increasing laser intensities. The grey shaded area indicates the time window in which the laser is active, while orange and green shaded areas correspond to the regions occupied by $MoS_2$ and $WSe_2$ layers, respectively. The grey dashed arrows highlight the distribution of peak amplitude in the vdW gap and vacuum region.

To confirm this mechanism, we evaluated the planar-averaged charge-density difference along the out-of-plane direction, $\Delta\rho(z) = \rho(z, t_{final}) - \rho(z, \mathrm{t_{initial}})$. In the weak-field regime ($10^{10}$ – $10^{11}$ $\mathrm{Wcm^{-2}}$), $\Delta\rho(z)$ exhibits primarily localized features within the individual layers (Figure 4b, shaded colored areas). However, at $5\mathrm{x}10^{11}$ $\mathrm{Wcm^{-2}}$, the density starts occupying the vdW gap, indicating the onset of out-of-plane redistribution. At $10^{12}$ $\mathrm{Wcm^{-2}}$, a pronounced feature develops above the $WSe_2$ layer, concomitant with an ~2.5-fold enhancement of the out-of-plane dipole relative to the ground state (Figure 4a). At the highest intensity ($I = 5\times10^{12}$ $\mathrm{Wcm^{-2}}$, black curve), the amplitude of the localized density oscillations within the $MoS_2$ (orange) region decreases substantially, whereas its redistribution into the vdW gap and outer vacuum regions becomes significantly more pronounced. This spatial redistribution accompanies the suppression of net interlayer charge transfer as a clear signature of Pauli blocking under strong field. This response remains robust under in-plane deformation (Figure S8), confirming strain as a non-invasive knob for fine-tuning the resonant conditions of the transient polarization. This extensive spatial expansion of $\Delta\rho(z)$ heavily dominates the dipole spatial integral $\Delta p_z = \int z\,\Delta\rho(z)dz$, generating a massive enhancement of the out-of-plane polarization, despite the saturation of net integrated charge transfer on the atomic planes.

To further examine state filling and its relation to Pauli blocking during interlayer charge transfer, we calculated the energy-resolved occupation density on the $MoS_2/WSe_2$ heterobilayer with equally distributed strain at two representative times: during the pulse (t = 8fs) and after the pulse (t = 30fs). In the weak-field regime ($10^{10}$ – $10^{11}$ $Wcm^{-2}$), the population of unoccupied states remains negligible at 1.1eV during irradiation (Figure 5a). After the pulse, the peak at 1.1. eV, corresponding to hybrid states sitting on both layers (Figure S1), gains weight due to the interlayer carrier relaxation. At intermediate intensity ($5\times10^{11}$ $Wcm^{-2}$), the population of these peak increases further, accompanied by the depletion of occupied states near -0.4 eV (Figure 5c). Upon increasing intensity ($1\times10^{12}$ $Wcm^{-2}$), the lowest-energy unoccupied states gain population by a factor of 1.5 during the pulse, with pronounced depopulation of occupied states in the [-0.5, -2.0 eV] range (Figure 5c). After the laser is turned off, carrier relaxation significantly populates these conduction states, reaching approximately twice their in-pulse amplitude (Figure 5d, magenta curve). Under strongest field (I = $5\times10^{12}$ $Wcm^{-2}$), the occupation profile during the pulse nearly doubles of the lowest unoccupied peak amplitude near 1.1 eV alongside significant spectral-weight transfer to higher unoccupied states ($> 2.0$ eV; Figure 5e). Notably, valence states down to -2 eV are almost completely depleted under this strong field. Once the field is turned off, two effects demonstrate state saturation and Pauli blocking: (i) in contrast to the $1\times10^{12}$ $Wcm^{-2}$ regime, where post-pulse relaxation further increases the lowest unoccupied peak at 1.1 eV, at I = $5\times10^{12}$ $Wcm^{-2}$ the post-pulse amplitude of this first peak decreases (compare Figures 5d and 5f); (ii) The excess of spectral weight is redistributed into higher-energy unoccupied states particularly in the 2.0 – 4.0 eV range (Figure 5 f). This redistribution is consistent with the saturation of the lowest states via Pauli blocking, which prevents further localized charge accumulation and redirects excited electronic density into higher-energy states and the vdW gap (see Figure 4b).

Table 1. Calculated interlayer charge-transfer density for different excitation regimes for $MoS_2/WSe_2$ heterobilayer with strain on both layers using averaged-charge per atom post-pulse.

| **Intensity regime** | **$\Delta q$/Atom ($10^{-3}$ e)** | **Carrier density ($n_{\mathrm{CT}}$, $cm^{-2}$)** |
|---|---|---|
| Weak ($1\times10^{10}$ $Wcm^{-2}$) | 8 | $2.6 \times 10^{13}$ |
| Medium ($1\text{-}5\times10^{11}$ $Wcm^{-2}$) | 20 | $6.6 \times 10^{13}$ |
| Strong ($1\text{-}5\times10^{12}$ $Wcm^{-2}$) | 30 | $8.9 \times 10^{13}$ |

Experimentally, low-field THz emission studies report charge-transfer saturation at fluences around 0.1 mJ $cm^{-2}$, where the buildup of a counter-field suppresses further charge separation [51]. In contrast, the stronger laser fields applied here overcome this internal opposition, enabling additional carrier redistribution. Under these non-perturbative conditions, charge accumulation within the atomic planes is instead limited by quantum Pauli blocking of localized conduction states. As localized states saturate, the strong driving field forces electronic density out of the atomic layers and into the vdW gap (Figure 4b), establishing a sub-15 fs transient polarization surge prior to the onset of incoherent thermalization and lattice motion [43,52,53].

These findings demonstrate the formation of a laser-induced metastable polarization across the $MoS_2/WSe_2$ heterobilayer. Through the interaction with a strong pulse, the system reaches a non-equilibrium state where the out-of-plane dipole moment is substantially enhanced by an unbalanced distribution of the electronic density across the interface. This induced polarization scales non-trivially with respect to the field intensity: while remaining substantially stable between $10^{11}$ and $10^{12}$ $Wcm^{-2}$, it increases dramatically at 5 x $10^{12}$ $Wcm^{-2}$. This behavior indicates that the transition to this polar state is not merely a perturbative response to absorption but is governed by a threshold-driven Pauli blocking mechanism. Entering the non-perturbative strong-field regime, state saturation suppresses localized interlayer charge accumulation, forcing the electronic density to delocalize into the interstitial vdW gap and vacuum regions, thereby maximizing the macroscopic out-of-plane dipole component.

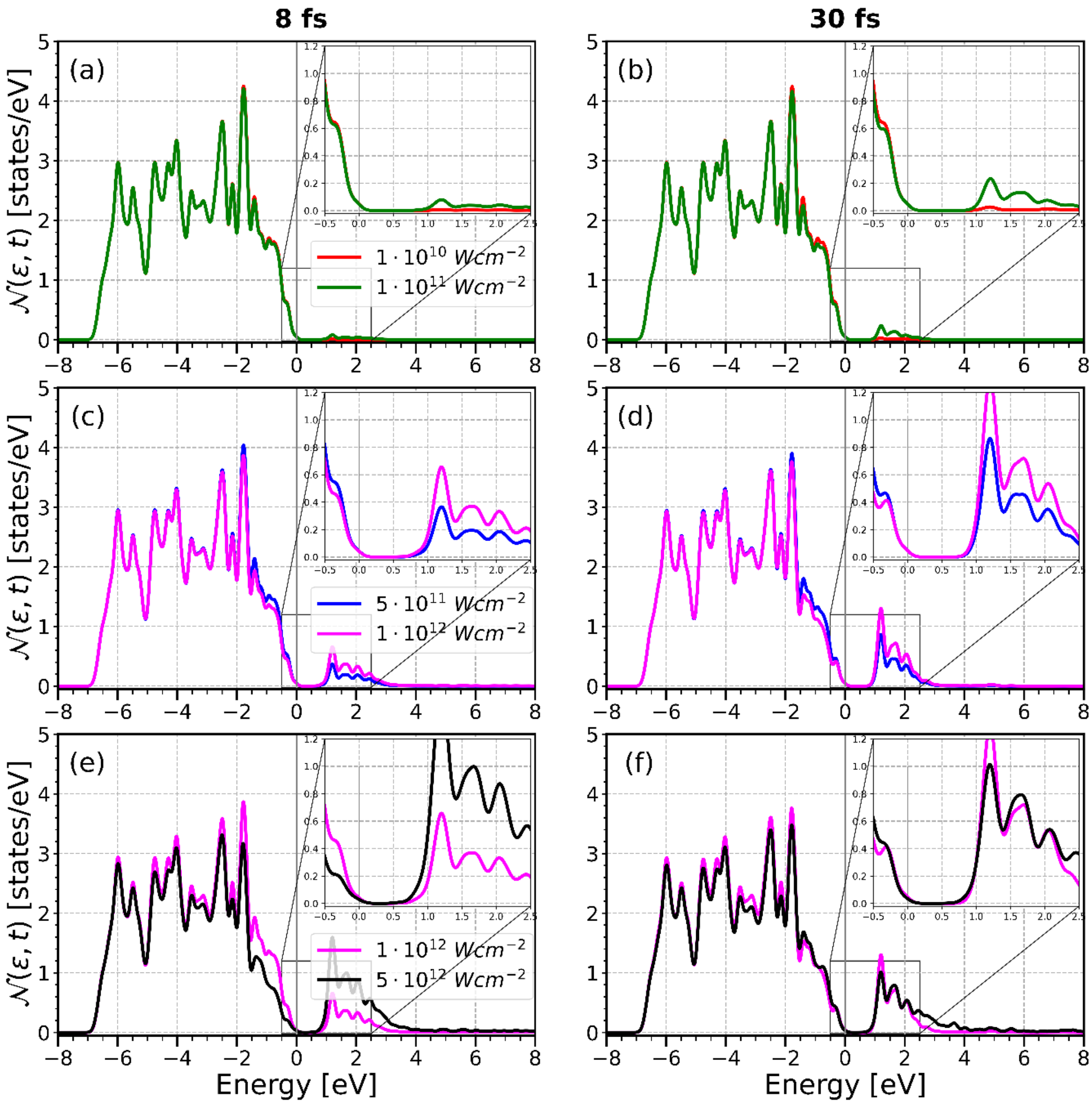

**Figure 5.** *Energy-resolved occupation density of the $MoS_2/WSe_2$ heterobilayer with strain on both layers under increasing laser intensities: 1×$10^{10}$ $Wcm^{-2}$ (red), 1×$10^{11}$ $Wcm^{-2}$ (green), 5×$10^{11}$ $Wcm^{-2}$ (blue), 1×$10^{12}$ $Wcm^{-2}$ (magenta), and 5×$10^{12}$ $Wcm^{-2}$ (black). The occupation densities are shown (a, c, and e; left panel) during the laser pulse (t = 8 fs) and (b, d, and f; right panel) after the pulse (t = 30 fs), while the inset magnifies the low-energy region near the Fermi level which is marked with grey line. An artificial broadening of 28 meV was applied to smoothen the curve.*

To establish a quantitative connection with pump-probe experiments, we evaluate the planar interlayer charge-transfer density $n_{\mathrm{CT}} = |\Delta q_{\mathrm{cell}}|/A_{\mathrm{cell}}$, where $\Delta q_{\mathrm{cell}}$ is the net Hirshfeld charge transferred per heterostructure unit cell ($A_{\mathrm{cell}} \approx 11.2$ A$^2$). As summarized in Table 1, across weak-to-intermediate field regimes ($10^{10} - 10^{11}$ W/cm$^2$), the calculated charge transfer corresponds to an excited carrier density scaling from $n_{\mathrm{CT}} \approx 2.6 \times 10^{13}$ cm$^{-2}$ up to $6.6 \times 10^{13}$ cm$^{-2}$. This quantitatively matches the photoexcited carrier densities reported in ultrafast transient reflection and optical pump-probe experiments on $MoS_2/WSe_2$ heterostructures ($10^{12} - 10^{13}$ cm$^{-2}$) [54,55].

While our RT-TDDFT simulations represent an idealized closed system in which the field-induced polarization persists after pulse termination due to unitary time evolution, its physical lifetime under experimental conditions will be governed by intrinsic electron-phonon scattering and energy dissipation [61,62]. Because the out-of-plane polarization surge is established within the sub-15 fs pulse envelope, it outruns typical electron-phonon relaxation (Table 2) by more than an order of magnitude [61,62,65], establishing a transient non-equilibrium state with an expected physical lifetime of ~50-100 fs prior to thermalization. Furthermore, while the high field intensities applied here ($10^{12} - 5 \times 10^{12}$ W/cm$^2$) drive non-perturbative dynamics, our ultrashort Gaussian pulse envelope ($\sigma = 2$ fs) restricts integrated fluences to between $5.0$ mJ/cm$^2$ and $25.1$ mJ/cm$^2$ (Table S1 and subsection S3). These values fall within or below the single-shot optical damage threshold for monolayer TMDs ($\sim 50 - 89$ mJ/cm$^2$) [66], confirming the experimental accessibility of this strong-field transient regime.

Table 2. Experimentally reported different relaxation timescales with their references.

| Dissipation method | Timescale | Ref. |
|---|---|---|
| Pulse-driven electronic polarization | 5 – 15 fs | preset work[56] |
| Carrier thermalization | 10 – 50 fs | [57] |
| Interlayer charge transfer | 50 – 100 fs | [58,59] |
| electron-phonon scattering | 0.1 – 10 ps | [61,62,63] |
| radiative recombination | 0 – 10 ns | [63,64] |

In conclusion, our findings establish that strong-field irradiation provides a robust route to engineer the femtosecond electronic response of TMD heterobilayers. By driving the system beyond the linear response regime, intense pulses unlock non-equilibrium states characterized by transiently emerging out-of-plane dipoles, driven by the spatial delocalization of electronic density into the interstitial vdW gap. Crucially, this mechanism remains remarkably effective across different in-plane strain configurations, demonstrating that the strong-field polarization response is an intrinsic, universal property of the interlayer potential. While our RT-TDDFT approach captures the instantaneous, sub-15 fs polarization surge during and immediately following laser irradiation, real-world applications will inevitably couple these dynamics to intrinsic and extrinsic relaxation pathways. Future theoretical investigations incorporating explicit electron-phonon scattering and decoherence channels will be essential to map the complete femtosecond-to-picosecond decay cascades of these transient states in open experimental environments. Overall, our results pave the way for sub-picosecond, all-optical control of interfacial electronic dipoles in next-generation light-wave-driven optoelectronic devices.

**Methods:**

The results reported in this work are obtained from first principles in the framework of real-time time-dependent density functional theory (TDDFT). In this formalism, the electronic dynamics are described by the time-dependent Kohn-Sham (TD-KS) equation [67],

$$i\hbar \frac{\partial}{\partial t}\psi_i(\boldsymbol{r},t) = \left(-\frac{\hbar^2}{2m}\nabla^2 + v_{\mathrm{eff}}(\boldsymbol{r},t)\right)\psi_i(\boldsymbol{r},t),$$

where $\psi_i(\boldsymbol{r},t)$ are the TD-KS orbitals building up the time-dependent electron density

$$n(\boldsymbol{r},t) = \sum_i^{occ} |\ \psi_i(\boldsymbol{r},t)\ |^2$$

The effective potential is given by

$$v_{\mathrm{eff}}\,(\mathbf{r},\mathrm{t}) = v_{\mathrm{ext}}(\mathbf{r},\mathrm{t}) + v_{\mathrm{H}}(\mathbf{r},\mathrm{t}) + v_{\mathrm{XC}}[n,\Psi_0](\mathbf{r},\mathrm{t}),$$

where $v_{\mathrm{ext}}$ is the time-dependent external potential, $v_{\mathrm{H}}$ is the Hartree potential, and $v_{\mathrm{XC}}$ is the exchange correlation potential, which is a functional of the time-dependent density $n(\boldsymbol{r},\boldsymbol{t})$ and the initial wavefunction $\Psi_0$.

The linear absorption spectrum is computed using the Yabana-Bertsch method [68], in which the time-dependent KS orbitals are propagated following a weak instantaneous phase-shift corresponding to a spatially uniform electric field,

$$\boldsymbol{E}(t) = E_0\delta(t).$$

Here, the external perturbation is applied only along the x-axis, taking advantage of the in-plane isotropy of the TMD heterostructure.

The interaction with a Gaussian pulse of varying intensity is simulated in the velocity-gauge, exploiting the known relation between the time-dependent electric field and the time-dependent vector potential:

$$\boldsymbol{E}(t) = -\frac{\partial \boldsymbol{A}(t)}{\partial t}.$$

The time-dependent laser intensity, $\mathrm{I}(t)$, is obtained from $\boldsymbol{E}(t)$ as:

$$\mathrm{I}(t) = \frac{c}{8\pi}|\boldsymbol{E}(t)|^2, \qquad I_{max} = max\big(\mathrm{I}(t)\big).$$

The energy fluence (F), defined as the incident laser energy per unit area, is obtained by integrating instantaneous intensity over pulse duration:

$$F = \int \mathrm{I}(t)\, dt \approx \sum_i I(t_i)\Delta t.$$

The corresponding fluence values and pulse-width definitions are summarized in Table S1, together with the relevant unit conversions.

The high-harmonic spectrum is calculated by Fourier-transforming the time derivative of the macroscopic current density

$$\boldsymbol{J}(t) = \frac{1}{\Omega}\int_{\Omega} \boldsymbol{j}(\boldsymbol{r},t)\, d^3r,$$

where $\Omega$ is the primitive unit cell volume and

$$\boldsymbol{j}(\boldsymbol{r},t) = \frac{1}{2}\sum_{\boldsymbol{ik}}\left[\phi^*_{\boldsymbol{ik}}(\boldsymbol{r},t)\left(-i\boldsymbol{\nabla} + \frac{\boldsymbol{A}(t)}{c}\right)\phi_{\boldsymbol{ik}}(\boldsymbol{r},t) + c.c.\right],$$

resulting in

$$HHG(\omega) = \int_0^\infty \omega^2 \mid \boldsymbol{J}(t)e^{i\omega t} \mid^2 dt.$$

The excitation energy per atom at time *t* is defined as:

$$E_{ex}(t) = \frac{\sum_{i,\boldsymbol{k}} \varepsilon_{\boldsymbol{ik}} f_{i,\boldsymbol{k}}(t)}{N_{\mathrm{atom}}},$$

where $\varepsilon_{\boldsymbol{ik}}$ are the ground-state KS energy eigenvalues, $f_{i,\boldsymbol{k}}(t)$ are the time-dependent populations, and $N_{\mathrm{atom}}$ is the number of atoms in the simulation box. The number of excited electrons per atom is given by

$$N_{ex}(t) = \frac{N_{\mathrm{el}} - \sum_{i,\boldsymbol{k}}^{\mathrm{occ}} f_{i,\boldsymbol{k}}(t)}{N_{\mathrm{atom}}} = \frac{\sum_{i,\boldsymbol{k}}^{\mathrm{unocc}} f_{i,\boldsymbol{k}}(t)}{N_{\mathrm{atom}}},$$

where $N_{\mathrm{el}}$ is the total number of electrons in the system at ground state. The energy uptake and number of excited electrons (in Figure 1) are determined by subtracting their ground state values (t = 0) from those at the end of simulation (t = 50fs). Their intensity dependence is analyzed by the power-law behavior,

$$Y\,(I) \propto I^{\alpha},$$

where *I* is the intensity and *Y* denotes the energy uptake or the number of excited electrons [69]. The scaling parameter α is obtained from a linear fit in the log-log data of both quantities.

All calculations are performed using real-space code Octopus [70] on a grid spacing of 0.158 Å with a 12×12×1 *k*-point mesh to sample the Brillouin zone. The crystal structures were initially optimized using the plane-wave code VASP [71] within the generalized-gradient approximation (PBE functional [72]) augmented by Grimme's DFT-D3 dispersion correction [73] to account for vdW interactions. A vacuum slab of 30 Å was used to separate periodic images in all calculations. For the strongest pulse intensity ($5 \times 10^{12}$ Wcm$^{-2}$), the vacuum thickness was further increased to 50Å to verify the adequacy of chosen vacuum spacing. The calculated time-dependent partial charges showed only negligible change, see Figure S8, confirming the accuracy of result using 30Å vacuum spacing. The linear absorption spectrum was obtained by applying an instantaneous kick k = 0.10 Å$^{-1}$ along the x direction[68]. In all calculations, Hartwigsen-Goedecker-Hutter pseudopotentials [74] are used. For the solution of the TD-KS equations, we employed the approximated enforced time-reversal symmetry propagator [75] with a time step of 2.41 attoseconds for a total simulation time of 50 fs. The interaction with a pulse was modeled using a Gaussian envelop with temporal standard deviation of σ = 2 fs (corresponding to a field-envelop FWHM of 3.33 fs), centered at $t_0$ = 8 fs and carrier frequency $\omega_0$ = 2.75 eV. Five peak intensities were considered: $1 \times 10^{10}$ Wcm$^{-2}$, $1 \times 10^{11}$ Wcm$^{-2}$, $5 \times 10^{11}$ Wcm$^{-2}$, $1 \times 10^{12}$ Wcm$^{-2}$, and $5 \times 10^{12}$ Wcm$^{-2}$.

**Supporting information:**

Structural and electronic properties; laser pulse parameters and fluence, Interlayer charge-transfer density; hierarchy of relaxation timescales; nonlinear optical response, including interlayer partial charges, high-harmonic generation spectra, out-of-plane dipole moment, and planar-averaged charge density along the z-axis for the strained variations; vacuum-thickness convergence test at the strongest laser field.

**Acknowledgements**

This work was funded by the German Research Foundation, project numbers 398816777 (CRC 1375, subprojects A8 and C4) and 547611111 (WHAT-A-TWIST).

**Data Availability Statement**

All data supporting the findings of this study have been deposited in Zenodo at002010.5281/zenodo.20625377. These data are publicly available free of charge.

**References**

[1] Frisk Kockum A, Miranowicz A, De Liberato S, Savasta S and Nori F 2019 Ultrastrong coupling between light and matter *Nat. Rev. Phys.* **1** 19–40

[2] Luo Y, Zhao J, Fieramosca A, Guo Q, Kang H, Liu X, Liew T C H, Sanvitto D, An Z, Ghosh S, Wang Z, Xu H and Xiong Q 2024 Strong light-matter coupling in van der Waals materials *Light Sci. Appl.* **13** 203

[3] Varjamo S-T, Edwards C, Zhou Y, Fang R, Hosseini Shokouh S H and Sun Z 2025 Optical Modification of TMD Heterostructures *Nano Lett.* **25** 4379–85

[4] Xu X, Lou Z, Cheng S, Chow P C Y, Koch N and Cheng H-M 2021 Van der Waals organic/inorganic heterostructures in the two-dimensional limit *Chem* **7** 2989–3026

[5] Roy R, Holec D, Kratzer M, Muenzer P, Kaushik P, Michal L, Kumar G S, Zajíčková L and Teichert C 2022 Probing the charge transfer and electron–hole asymmetry in graphene–graphene quantum dot heterostructure *Nanotechnology* **33** 325704

[6] Liew T C H, Kavokin A V and Shelykh I A 2008 Optical Circuits Based on Polariton Neurons in Semiconductor Microcavities *Phys. Rev. Lett.* **101** 016402

[7] Ballarini D, De Giorgi M, Cancellieri E, Houdré R, Giacobino E, Cingolani R, Bramati A, Gigli G and Sanvitto D 2013 All-optical polariton transistor *Nat. Commun.* **4** 1778

[8] Guo X, Lyu W, Chen T, Luo Y, Wu C, Yang B, Sun Z, García de Abajo F J, Yang X and Dai Q 2023 Polaritons in Van der Waals Heterostructures *Adv. Mater.* **35** 2201856

[9] Kipp G, Bretscher H M, Schulte B, Herrmann D, Kusyak K, Day M W, Kesavan S, Matsuyama T, Li X, Langner S M, Hagelstein J, Sturm F, Potts A M, Eckhardt C J, Huang Y, Watanabe K, Taniguchi T, Rubio A, Kennes D M, Sentef M A, Baudin E, Meier G, Michael M H and McIver J W 2025 Cavity electrodynamics of van der Waals heterostructures *Nat. Phys.* **21** 1926–33

[10] Klimmer S, Ghaebi O, Gan Z, George A, Turchanin A, Cerullo G and Soavi G 2021 All-optical polarization and amplitude modulation of second-harmonic generation in atomically thin semiconductors *Nat. Photonics* **15** 837–42

[11] Herrmann P, Klimmer S, Lettau T, Monfared M, Staude I, Paradisanos I, Peschel U and Soavi G 2023 Nonlinear All-Optical Coherent Generation and Read-Out of Valleys in Atomically Thin Semiconductors *Small* **19** 2301126

[12] Dogadov O, Trovatello C, Yao B, Soavi G and Cerullo G 2022 Parametric Nonlinear Optics with Layered Materials and Related Heterostructures *Laser Photonics Rev.* **16** 2100726

[13] Friedrich F, Herrmann P, Shanbhag S S, Klimmer S, Wilhelm J and Soavi G 2026 Measurement of optically induced broken time-reversal symmetry in atomically thin crystals *Nat. Photonics* **20** 186–93

[14] Zasedatelev A V, Baranikov A V, Urbonas D, Scafirimuto F, Scherf U, Stöferle T, Mahrt R F and Lagoudakis P G 2019 A room-temperature organic polariton transistor *Nat. Photonics* **13** 378–83

[15] Sie E J, Nyby C M, Pemmaraju C D, Park S J, Shen X, Yang J, Hoffmann M C, Ofori-Okai B K, Li R, Reid A H, Weathersby S, Mannebach E, Finney N, Rhodes D, Chenet D, Antony A, Balicas L, Hone J, Devereaux T P, Heinz T F, Wang X and Lindenberg A M 2019 An ultrafast symmetry switch in a Weyl semimetal *Nature* **565** 61–6

[16] Johnson A S, Pastor E, Batlle-Porro S, Benzidi H, Katayama T, de la Peña Muñoz G A, Krapivin V, Kim S, López N, Trigo M and Wall S E 2024 All-optical seeding of a light-induced phase transition with correlated disorder *Nat. Phys.* **20** 970–5

[17] Hervé M, Privault G, Trzop E, Akagi S, Watier Y, Zerdane S, Chaban I, Torres Ramírez R G, Mariette C, Volte A, Cammarata M, Levantino M, Tokoro H, Ohkoshi S and Collet E 2024 Ultrafast and persistent photoinduced phase transition at room temperature monitored by streaming powder diffraction *Nat. Commun.* **15** 267

[18] Wada O 2004 Femtosecond all-optical devices for ultrafast communication and signal processing *New J. Phys.* **6** 183

[19] Amo A, Liew T C H, Adrados C, Houdré R, Giacobino E, Kavokin A V and Bramati A 2010 Exciton–polariton spin switches *Nat. Photonics* **4** 361–6

[20] Krumland J and Cocchi C 2021 Conditions for electronic hybridization between transition-metal dichalcogenide monolayers and physisorbed carbon-conjugated molecules *Electron. Struct.* **3** 044003

[21] Di Giorgio C, Blundo E, Pettinari G, Felici M, Bobba F and Polimeni A 2022 Mechanical, Elastic, and Adhesive Properties of Two-Dimensional Materials: From Straining Techniques to State-of-the-Art Local Probe Measurements *Adv. Mater. Interfaces* **9** 2102220

[22] Stellino E, D'Alò B, Blundo E, Postorino P and Polimeni A 2024 Fine-Tuning of the Excitonic Response in Monolayer WS2 Domes via Coupled Pressure and Strain Variation *Nano Lett.* **24** 3945–51

[23] Ramasubramaniam A 2012 Large excitonic effects in monolayers of molybdenum and tungsten dichalcogenides *Phys. Rev. B* **86** 115409

[24] Ross J S, Klement P, Jones A M, Ghimire N J, Yan J, Mandrus D G, Taniguchi T, Watanabe K, Kitamura K, Yao W, Cobden D H and Xu X 2014 Electrically tunable excitonic light-emitting diodes based on monolayer WSe2 p–n junctions *Nat. Nanotechnol.* **9** 268–72

[25] Chernikov A, van der Zande A M, Hill H M, Rigosi A F, Velauthapillai A, Hone J and Heinz T F 2015 Electrical Tuning of Exciton Binding Energies in Monolayer ${\mathrm{WS}}_{2}$ *Phys. Rev. Lett.* **115** 126802

[26] Barbone M, Montblanch A R-P, Kara D M, Palacios-Berraquero C, Cadore A R, De Fazio D, Pingault B, Mostaani E, Li H, Chen B, Watanabe K, Taniguchi T, Tongay S, Wang G, Ferrari A C

and Atatüre M 2018 Charge-tuneable biexciton complexes in monolayer WSe2 *Nat. Commun.* **9** 3721

[27] Wu R, Zhang H, Ma H, Zhao B, Li W, Chen Y, Liu J, Liang J, Qin Q, Qi W, Chen L, Li J, Li B and Duan X 2024 Synthesis, Modulation, and Application of Two-Dimensional TMD Heterostructures *Chem. Rev.* **124** 10112–91

[28] Geim A K and Grigorieva I V 2013 Van der Waals heterostructures *Nature* **499** 419–25

[29] Kunstmann J, Mooshammer F, Nagler P, Chaves A, Stein F, Paradiso N, Plechinger G, Strunk C, Schüller C, Seifert G, Reichman D R and Korn T 2018 Momentum-space indirect interlayer excitons in transition metal dichalcogenide van der Waals heterostructures *Nat. Phys.* **14** 801–5

[30] Ramzan M S, Kunstmann J and Kuc A B 2021 Tuning Valleys and Wave Functions of van der Waals Heterostructures by Varying the Number of Layers: A First-Principles Study *Small* 2008153

[31] Yuan L, Zheng B, Kunstmann J, Brumme T, Kuc A B, Ma C, Deng S, Blach D, Pan A and Huang L 2020 Twist-angle-dependent interlayer exciton diffusion in WS2–WSe2 heterobilayers *Nat. Mater.* **19** 617–23

[32] Zhu X, Mocatti S and Calandra M 2026 Light-Induced Transient Polarization Reversal in Rhombohedrally Stacked Bilayer Transition Metal Dichalcogenides via an Electronic Mechanism *Phys. Rev. Lett.* **137** 106402

[33] Zhang C, Li M-Y, Tersoff J, Han Y, Su Y, Li L-J, Muller D A and Shih C-K 2018 Strain distributions and their influence on electronic structures of WSe2–MoS2 laterally strained heterojunctions *Nat. Nanotechnol.* **13** 152–8

[34] Hong S C, Baek J-H, Chang Y, Nguyen M H, Park W, Lim C, Kim J, Lee H, Kim C, Watanabe K, Taniguchi T, Son J, Han J W, Cheong H, Kim M and Lee G-H 2025 Overcoming the Lattice Mismatch Barrier for Atomic Reconstruction in MoSe2/MoS2 Heterobilayers *ACS Nano* **19** 41233–43

[35] Ghaebi O, Hamzayev T, Weickhardt T, Ramzan M S, Taniguchi T, Watanabe K, Cocchi C, De Fazio D and Soavi G 2025 Tunable Exciton Modulation and Efficient Charge Transfer in $MoS_2$ /Graphene van der Waals Heterostructures *ACS Nano* **19** 19027–34

[36] Ramzan M S and Cocchi C 2025 Strain-Engineered Level Alignment in the $MoTe_2$ /$WSe_2$ Heterobilayer *Phys. Status Solidi RRL – Rapid Res. Lett.* **19** 2400276

[37] Peng Z, Chen X, Fan Y, Srolovitz D J and Lei D 2020 Strain engineering of 2D semiconductors and graphene: from strain fields to band-structure tuning and photonic applications *Light Sci. Appl.* **9** 190

[38] Cho C, Wong J, Taqieddin A, Biswas S, Aluru N R, Nam S and Atwater H A 2021 Highly Strain-Tunable Interlayer Excitons in MoS2/WSe2 Heterobilayers *Nano Lett.* **21** 3956–64

[39] Hussain M, Ghaebi O, Monfared M, Gruenewald M, Ahsan U, Lipilin F, Luxa J, Sofer Z, Peschel U and Soavi G 2025 Nonlinear Optical Properties of Mono and Multilayer MoWSe2 Alloys *Adv. Opt. Mater.* **13** e01000

[40] Liang S, Lu Y, Liu H, Shang X, Du R, Ji J, Yu Y and Zhang S 2025 Saturable absorption of few-layer $WS_2$ and $WSe_2$ at exciton resonance *Opt. Express* **33** 7266–77

[41] Liu H, Li Y, You Y S, Ghimire S, Heinz T F and Reis D A 2017 High-harmonic generation from an atomically thin semiconductor *Nat. Phys.* **13** 262–5

[42] Wang Y, Iyikanat F, Bai X, Hu X, Das S, Dai Y, Zhang Y, Du L, Li S, Lipsanen H, García de Abajo F J and Sun Z 2022 Optical Control of High-Harmonic Generation at the Atomic Thickness *Nano Lett.* **22** 8455–62

[43] Jacobs M, Krumland J, Valencia A M, Wang H, Rossi M and Cocchi C 2020 Ultrafast charge transfer and vibronic coupling in a laser-excited hybrid inorganic/organic interface *Adv. Phys. X* **5** 1749883

[44] Jacobs M, Krumland J and Cocchi C 2022 Laser-Controlled Charge Transfer in a Two-Dimensional Organic/Inorganic Optical Coherent Nanojunction *ACS Appl. Nano Mater.* **5** 5187–95

[45] Chen Z, Zhu M, Ren T, He J, Loh K P and Xu Q-H 2021 Transient Reflection Spectroscopy on Ultrafast Interlayer Charge Transfer Processes in a MoS2/WSe2 van der Waals Heterojunction *J. Phys. Chem. C* **125** 26575–82

[46] Choi J, Embley J, Blach D D, Perea-Causín R, Erkensten D, Kim D S, Yuan L, Yoon W Y, Taniguchi T, Watanabe K, Ueno K, Tutuc E, Brem S, Malic E, Li X and Huang L 2023 Fermi Pressure and Coulomb Repulsion Driven Rapid Hot Plasma Expansion in a van der Waals Heterostructure *Nano Lett.* **23** 4399–405

[47] Xu C, Barden N, Alexeev E M, Wang X, Long R, Cadore A R, Paradisanos I, Ott A K, Soavi G, Tongay S, Cerullo G, Ferrari A C, Prezhdo O V and Loh Z-H 2024 Ultrafast Charge Transfer and Recombination Dynamics in Monolayer–Multilayer WSe2 Junctions Revealed by Time-Resolved Photoemission Electron Microscopy *ACS Nano* **18** 1931–47

[48] Wu Y, Duan J, Ma W, Ou Q, Li P, Alonso-González P, Caldwell J D and Bao Q 2022 Manipulating polaritons at the extreme scale in van der Waals materials *Nat. Rev. Phys.* **4** 578–94

[49] Datta B, Khatoniar M, Deshmukh P, Thouin F, Bushati R, De Liberato S, Cohen S K and Menon V M 2022 Highly nonlinear dipolar exciton-polaritons in bilayer MoS2 *Nat. Commun.* **13** 6341

[50] Louca C, Genco A, Chiavazzo S, Lyons T P, Randerson S, Trovatello C, Claronino P, Jayaprakash R, Hu X, Howarth J, Watanabe K, Taniguchi T, Dal Conte S, Gorbachev R, Lidzey D G, Cerullo G, Kyriienko O and Tartakovskii A I 2023 Interspecies exciton interactions lead to enhanced nonlinearity of dipolar excitons and polaritons in MoS2 homobilayers *Nat. Commun.* **14** 3818

[51] Ma E Y, Guzelturk B, Li G, Cao L, Shen Z-X, Lindenberg A M and Heinz T F 2019 Recording interfacial currents on the subnanometer length and femtosecond time scale by terahertz emission *Sci. Adv.* **5** eaau0073

[52] Jacobs M, Fidanyan K, Rossi M and Cocchi C 2024 Impact of nuclear effects on the ultrafast dynamics of an organic/inorganic mixed-dimensional interface *Electron. Struct.* **6** 025006

[53] Souri S, Timmer D, Lünemann D C, Hadilou N, Winte K, De Sio A, Esmann M, Curdt F, Winklhofer M, Anhäuser S, Guerrini M, Valencia A M, Cocchi C, Witte G and Lienau C 2024 Ultrafast Time-Domain Spectroscopy Reveals Coherent Vibronic Couplings upon Electronic Excitation in Crystalline Organic Thin Films *J. Phys. Chem. Lett.* **15** 11170–81

[54] Hong X, Kim J, Shi S F, Zhang Y, Jin C, Sun Y, Tongay S, Wu J, Zhang Y and Wang F 2014 Ultrafast charge transfer in atomically thin MoS2/WS2 heterostructures *Nat. Nanotechnol.* **9** 682–6

[55] Chen H, Wen X, Zhang J, Wu T, Gong Y, Zhang X, Yuan J, Yi C, Lou J, Ajayan P M, Zhuang W, Zhang G and Zheng J 2016 Ultrafast formation of interlayer hot excitons in atomically thin MoS2/WS2 heterostructures *Nat. Commun.* **7** 12512

[56] Ramzan M S, Soavi G and Cocchi C 2026 All-Optical Control of Interfacial Polarization in MoS$_2$/WSe$_2$ Heterobilayers

[57] Nie Z, Long R, Sun L, Huang C-C, Zhang J, Xiong Q, Hewak D W, Shen Z, Prezhdo O V and Loh Z-H 2014 Ultrafast Carrier Thermalization and Cooling Dynamics in Few-Layer MoS2 *ACS Nano* **8** 10931–40

[58] Jiang Y, Chen S, Zheng W, Zheng B and Pan A 2021 Interlayer exciton formation, relaxation, and transport in TMD van der Waals heterostructures *Light Sci. Appl.* **10** 72

[59] Policht V R, Mittenzwey H, Dogadov O, Katzer M, Villa A, Li Q, Kaiser B, Ross A M, Scotognella F, Zhu X, Knorr A, Selig M, Cerullo G and Dal Conte S 2023 Time-domain observation of interlayer exciton formation and thermalization in a MoSe2/WSe2 heterostructure *Nat. Commun.* **14** 7273

[60] Chi Z, Chen H, Chen Z, Zhao Q, Chen H and Weng Y-X 2018 Ultrafast Energy Dissipation via Coupling with Internal and External Phonons in Two-Dimensional MoS2 *ACS Nano* **12** 8961–9

[61] Jin C, Ma E Y, Karni O, Regan E C, Wang F and Heinz T F 2018 Ultrafast dynamics in van der Waals heterostructures *Nat. Nanotechnol.* **13** 994–1003

[62] Richter M 2024 Theory of interlayer exciton dynamics in two-dimensional transition metal dichalcogenide heterolayers under the influence of strain reconstruction and disorder *Phys. Rev. B* **109** 125308

[63] Palummo M, Bernardi M and Grossman J C 2015 Exciton radiative lifetimes in two-dimensional transition metal dichalcogenides *Nano Lett.* **15** 2794–800

[64] Barré E, Karni O, Liu E, O'Beirne A L, Chen X, Ribeiro H B, Yu L, Kim B, Watanabe K, Taniguchi T, Barmak K, Lui C H, Refaely-Abramson S, da Jornada F H and Heinz T F 2022 Optical absorption of interlayer excitons in transition-metal dichalcogenide heterostructures *Science* **376** 406–10

[65] Ji Z, Hong H, Zhang J, Zhang Q, Huang W, Cao T, Qiao R, Liu C, Liang J, Jin C, Jiao L, Shi K, Meng S and Liu K 2017 Robust Stacking-Independent Ultrafast Charge Transfer in MoS2/WS2 Bilayers *ACS Nano* **11** 12020–6

[66] Paradisanos I, Kymakis E, Fotakis C, Kioseoglou G and Stratakis E 2014 Intense femtosecond photoexcitation of bulk and monolayer MoS2 *Appl. Phys. Lett.* **105** 041108

[67] Runge E and Gross E K U 1984 Density-Functional Theory for Time-Dependent Systems *Phys. Rev. Lett.* **52** 997–1000

[68] Yabana K and Bertsch G F 1996 Time-dependent local-density approximation in real time *Phys. Rev. B* **54** 4484–7

[69] Bellersen H, Guerrini M and Cocchi C 2025 Ultrafast photoexcitation of semiconducting photocathode materials *Phys. Rev. B* **112** 024314

[70] Tancogne-Dejean N, Oliveira M J T, Andrade X, Appel H, Borca C H, Le Breton G, Buchholz F, Castro A, Corni S, Correa A A, De Giovannini U, Delgado A, Eich F G, Flick J, Gil G, Gomez A, Helbig N, Hübener H, Jestädt R, Jornet-Somoza J, Larsen A H, Lebedeva I V, Lüders M, Marques M A L, Ohlmann S T, Pipolo S, Rampp M, Rozzi C A, Strubbe D A, Sato S A, Schäfer C, Theophilou I, Welden A and Rubio A 2020 Octopus, a computational framework for exploring light-driven phenomena and quantum dynamics in extended and finite systems *J. Chem. Phys.* **152** 124119

[71] Kresse G and Furthmüller J 1996 Efficient iterative schemes for *ab initio* total-energy calculations using a plane-wave basis set *Phys. Rev. B* **54** 11169–86

[72] Perdew J P, Burke K and Ernzerhof M 1996 Generalized Gradient Approximation Made Simple *Phys. Rev. Lett.* **77** 3865–8

[73] Grimme S, Antony J, Ehrlich S and Krieg H 2010 A consistent and accurate ab initio parametrization of density functional dispersion correction (DFT-D) for the 94 elements H-Pu *J. Chem. Phys.* **132** 154104

[74] Hartwigsen C, Goedecker S and Hutter J 1998 Relativistic separable dual-space Gaussian pseudopotentials from H to Rn *Phys. Rev. B* **58** 3641–62

[75] Castro A, Marques M A L and Rubio A 2004 Propagators for the time-dependent Kohn–Sham equations *J. Chem. Phys.* **121** 3425–33

*Supporting Information*

# All-Optical Control of Interfacial Polarization in $MoS_2/WSe_2$ Heterobilayers

Muhammad Sufyan Ramzan,[1] Giancarlo Soavi,[2,3] and Caterina Cocchi[1,3]

[1] *Institut für Festkörpertheorie und Optik, Friedrich-Schiller-Universität Jena, 07743 Jena, Germany*

[2]*Institut für Festkörperphysik,*[2]*Friedrich-Schiller-Universität Jena, 07743 Jena, Germany*

[3]*Abbe Center of Photonics, Friedrich-Schiller-Universität Jena, 07745 Jena, Germany*

# S1. Structural properties

The optimized in-plane lattice parameters for monolayer (ML) $MoS_2$ and $WSe_2$ are 3.166 Å and 3.294 Å, respectively. For the $MoS_2/WSe_2$ heterobilayer with equally distributed strain, the converged lattice parameter is 3.226 Å, which corresponds to a 1.90% tensile strain in the $MoS_2$ layer and a -2.06%compressive strain in $WSe_2$. In the configuration where strain is applied to $MoS_2$ (referred to as $\varepsilon_{MoS2}$ in the main text), the lattice parameter of the heterobilayer is fixed to match that of ML $WSe_2$ (3.294 Å). The atomic positions are relaxed until the forces and total energies meet the convergence criteria. An analogous procedure is adopted for the configuration where the strain is localized entirely within the $WSe_2$ layer ($\varepsilon_{WSe2}$), where the lattice parameter of the heterobilayer is fixed to that of pristine ML $MoS_2$ (3.166 Å).

# S2. Electronic properties

The $MoS_2/WSe_2$ heterobilayer exhibits a type-II band alignment, with the valence band maximum (VBM) localized on ML $WSe_2$ and the conduction band minimum (CBM) on $MoS_2$ (Figure S1). To validate our computational parameters and real-space grid choices, we reproduced the electronic structure of the $MoS_2/WSe_2$ heterobilayer under various strain configurations (Figure S1). The obtained band structures are in good agreement with corresponding VASP calculations, confirming the reliability of the real-space Octopus simulations.

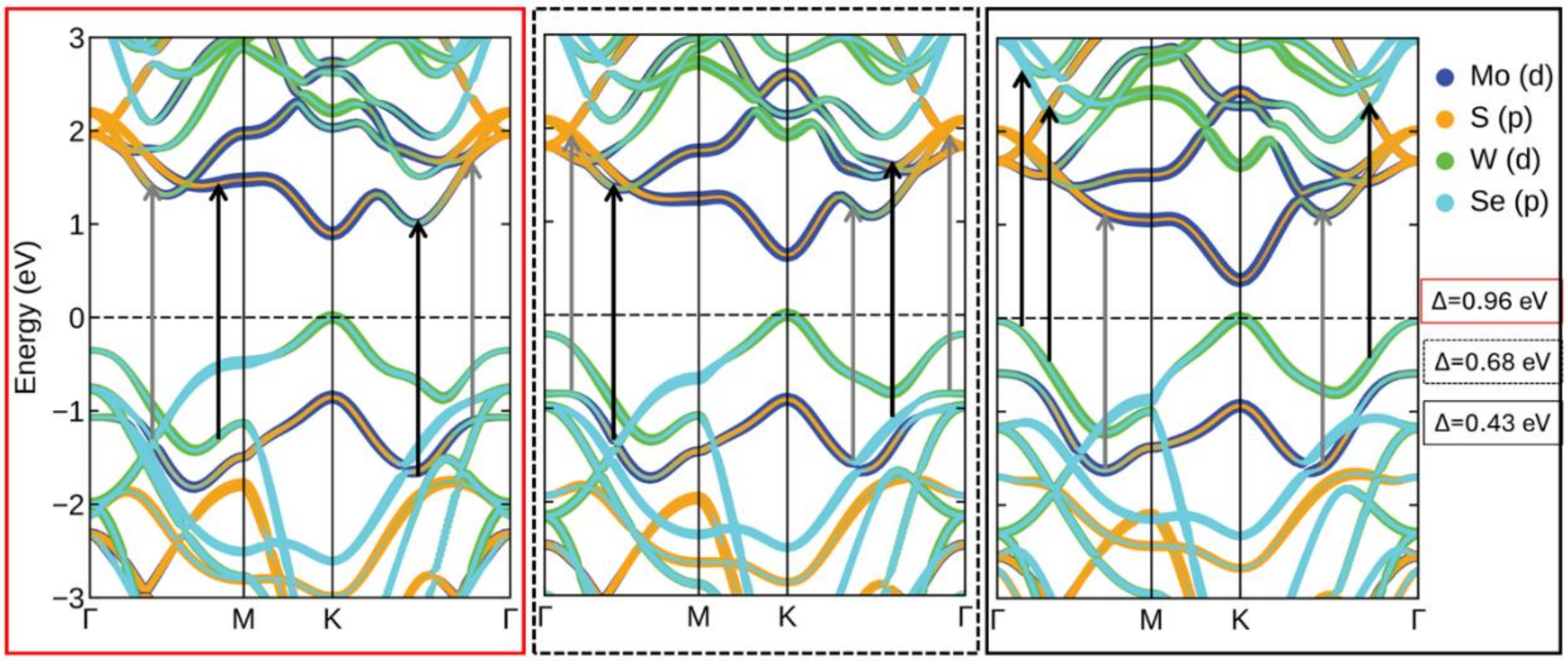

Figure S 1. Atom-projected electronic band structure of $MoS_2/WSe_2$ heterobilayer (atomic orbital contributions color-coded according to the legend) under in-plane strain applied entirely to $WSe_2$ (red solid box) and $MoS_2$ (black solid box) and equally distributed between both layers (black dashed box). These calculations were performed with VASP, using the PBE functional and neglecting spin-orbit coupling. The Fermi-level is set to 0 eV and the fundamental band gaps Δ of each configuration are reported in the legend. Black and grey arrows indicate the electronic transitions contributing with large (>10%) and moderate (<10%) weight to the primary optical resonance at 2.75 eV. It is worth noting that in all three configurations there is no contribution coming from direct transitions at K.

# S3. Laser pulse parameters and Fluence

The laser field is introduced through a time-dependent vector potential of the form

$$A(t) = A_0\, g(t) \cos\, [\omega(t - t_0)] \quad (S1)$$

here $A_0$is the vector-potential amplitude, $\omega$ is the carrier frequency, $t_0$is the pulse center (in our simulations $t_0 = \ 8fs$), and the Gaussian envelope, $g(t)$, is described as:

$$g(t) = \exp\left[-\frac{(t - t_0)^2}{2\tau_0^2}\right] \quad (S2)$$

The corresponding electric field $E(t)$, the instantaneous laser intensity, and the peak intensity is calculated as:

$$E(t) = -\frac{dA(t)}{dt} \qquad (S3)$$

$$I(t) = \frac{c}{8\pi}|E(t)|^2 \quad (S4)$$

The energy fluence (F), defined as the incident laser energy per unit area, is obtained by integrating instantaneous intensity over pulse time:

$$F = \int I\,(t)\,dt \approx \sum_i I\,(t_i)\Delta t \qquad (S5)$$

where $t_i$ denotes the $i$-th time step and $\Delta t$ is the temporal resolution of the simulation (2.41 as). Thus, the fluence accounts for the complete temporal profile of the pulse rather than being estimated simply as $I_{\text{peak}}$multiplied by a nominal pulse duration.

It is worth highlighting that the full width at half maximum (FWHM) of laser incorporated in the Octopus simulations is narrower due to squaring of envelope (Eq. S4). Therefore, the Gaussian envelope used here, $\tau_0 = 2$ fs and $t_0 = 8$ fs would have a FWHM as:

$$\text{FWHM}_A = 2\sqrt{2\ln 2}\ \tau_0 = \ 4.71\ fs,$$

while the FWHM of the corresponding squared Gaussian envelops is

$$\text{FWHM}_I = 2\sqrt{\ln 2}\ \tau_0 = \ 3.33\ fs.$$

Table S1. Time-integrated intensity (Int. Intensity) and corresponding fluence reported by the Octopus code in atomic units for each laser intensity, together with their converted values in physical units.

| Peak intensity (W/cm²) | Int. Intensity (a.u.) | Fluence (a.u.) | Fluence (mJ/cm²) |
|---|---|---|---|
| $1.0 \times 10^{10}$ | $1.149 \times 10^{-4}$ | $2.110 \times 10^{-5}$ | 0.018 |
| $1.0 \times 10^{11}$ | $1.149 \times 10^{-3}$ | $2.110 \times 10^{-4}$ | 0.179 |
| $5.0 \times 10^{11}$ | $5.750 \times 10^{-3}$ | $1.055 \times 10^{-3}$ | 0.895 |
| $1.0 \times 10^{12}$ | $1.150 \times 10^{-2}$ | $2.109 \times 10^{-3}$ | 1.790 |
| $5.0 \times 10^{12}$ | $5.750 \times 10^{-2}$ | $1.055 \times 10^{-2}$ | 8.950 |

# S4. Nonlinear optical response

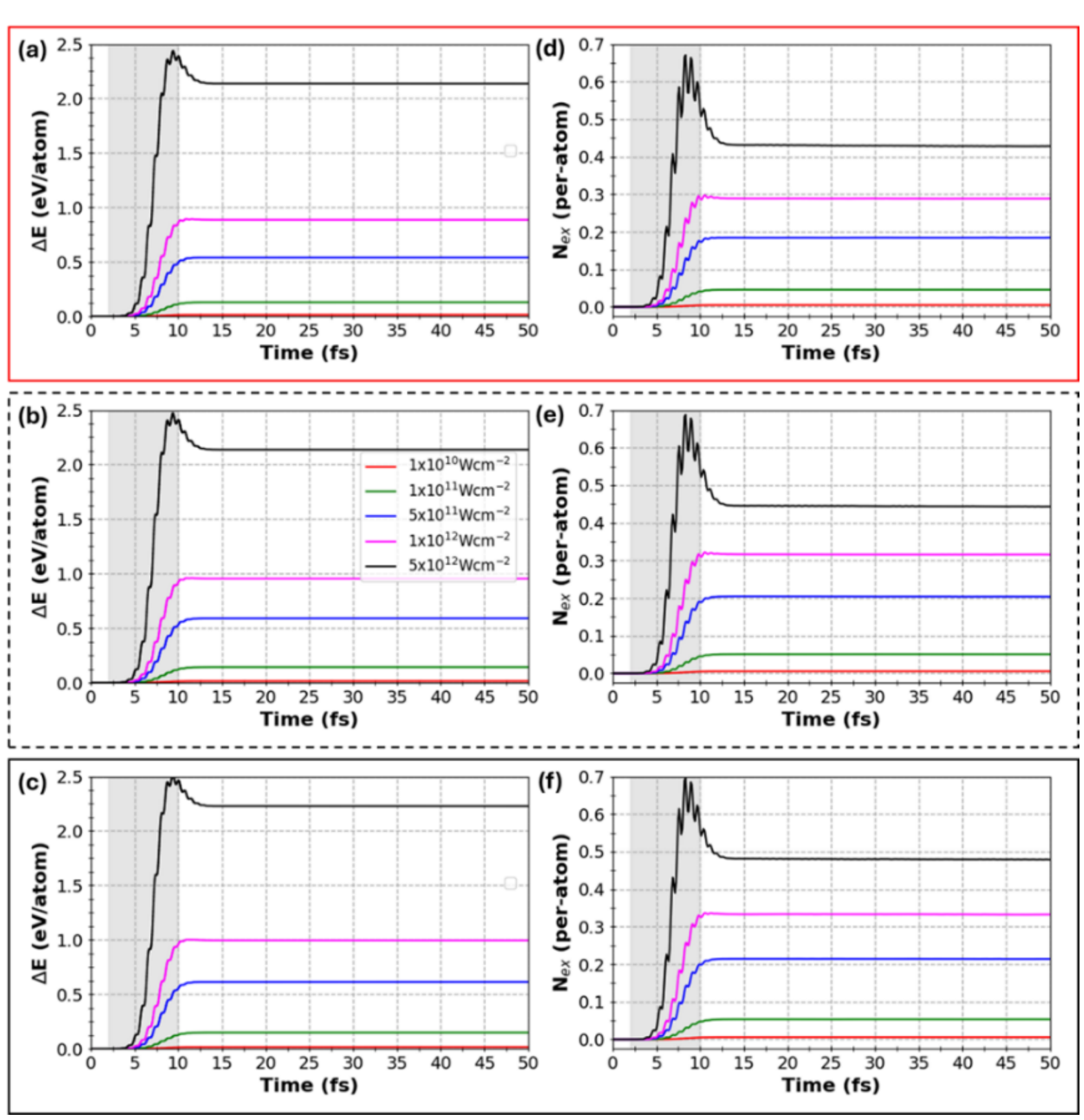

Figure S 2. (a) - (c) Time evolution of the electronic energy uptake (ΔE) and (d) - (f) number of excited electron per atom ($N_{ex}$) under varying laser intensities corresponding to $1\times10^{10}$ Wcm$^{-2}$ (red), $1\times10^{11}$ Wcm$^{-2}$ (green), $5\times10^{11}$ Wcm$^{-2}$ (blue), $1\times10^{12}$ Wcm$^{-2}$ (magenta), and $5\times10^{12}$ Wcm$^{-2}$ (black). Results obtained for the $MoS_2/WSe_2$ heterobilayer with in-plane strain applied to $WSe_2$ and $MoS_2$ monolayers are reported in the red and black solid boxes, respectively, while those computed with equally distributed strain are in the black dashed box. The grey shaded regions mark the window of laser irradiation.

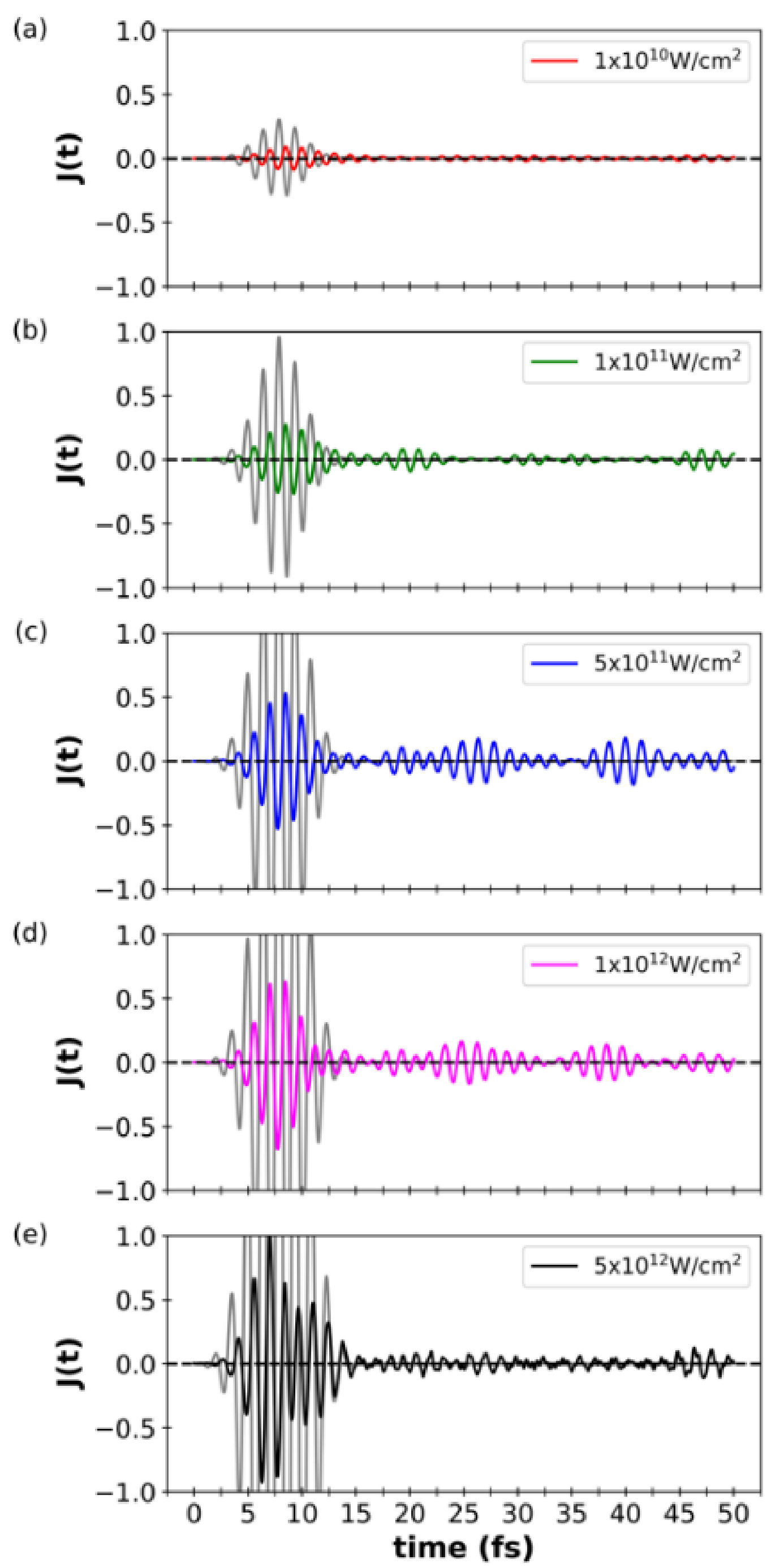

Figure S 3. Time evolution of current densities J(t) in $MoS_2/WSe_2$ heterobilayer with average strain, under laser pulses of increasing peak intensities: (a) $1\times10^{10}$ $Wcm^{-2}$, (b) $1\times10^{11}$ $Wcm^{-2}$, (c) $5\times10^{11}$ $Wcm^{-2}$, (d) $1\times10^{12}$ $Wcm^{-2}$, and (e) $5\times10^{12}$ $Wcm^{-2}$, shown in grey in the background.

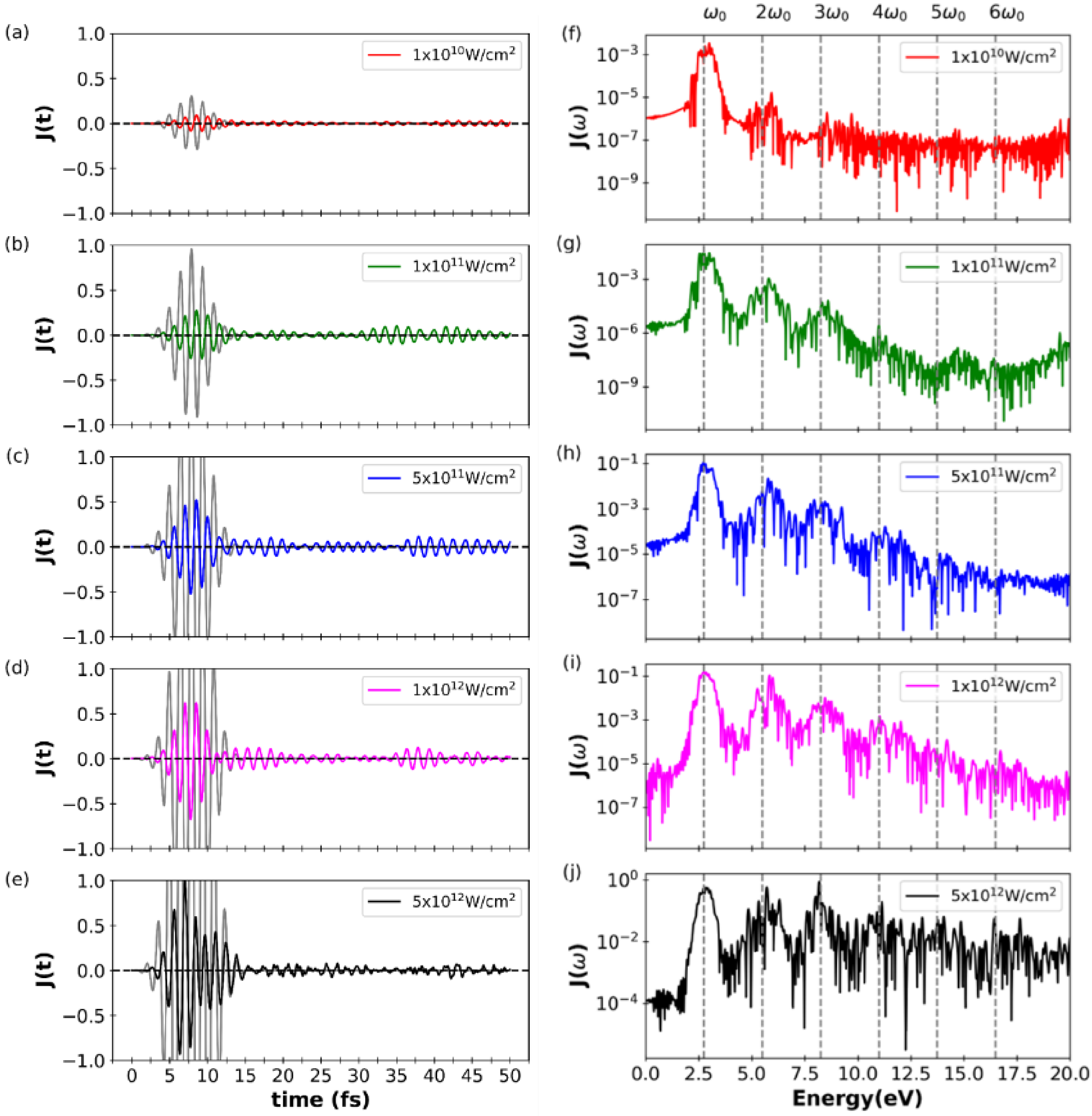

Figure S 4. Time evolution of current densities J(t) in $MoS_2/WSe_2$ heterobilayer with strain applied exclusively to the $MoS_2$ ML, under laser pulses of increasing peak intensities: (a) $1\times10^{10}$ $Wcm^{-2}$, (b) $1\times10^{11}$ $Wcm^{-2}$, (c) $5\times10^{11}$ $Wcm^{-2}$, (d) $1\times10^{12}$ $Wcm^{-2}$, and (e) $5\times10^{12}$ $Wcm^{-2}$, shown in grey in the background. (f) – (j) Corresponding high-harmonic spectra J(ω), obtained via Fourier transformation of the induced current densities in (a-e). The fundamental carrier frequency ($\omega_0$) and its harmonics are indicated by vertical dashed lines.

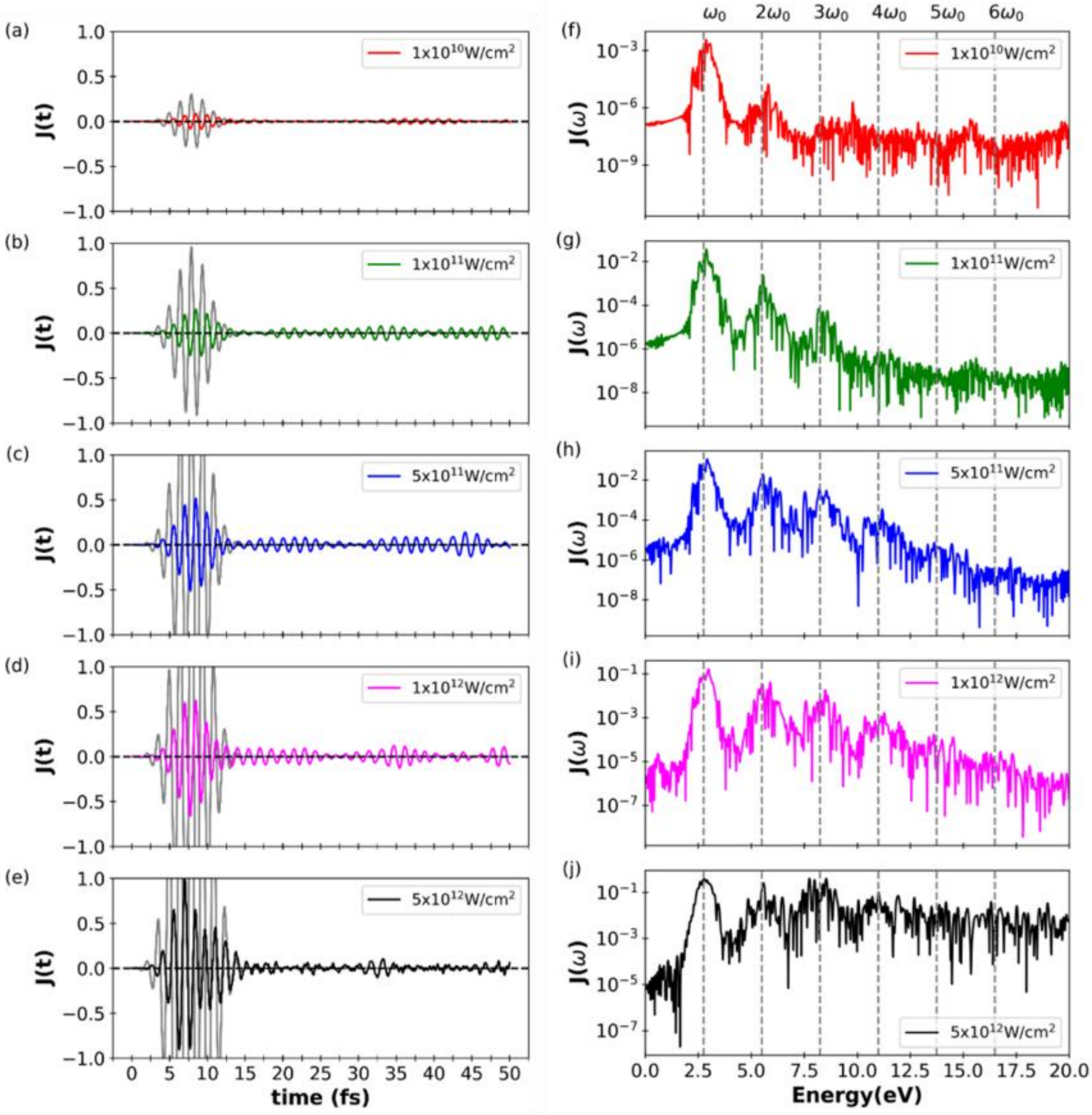

Figure S 5. Time evolution of current densities J(t) in $MoS_2/WSe_2$ heterobilayer with strain applied exclusively to the $WSe_2$ ML, under laser pulses of increasing peak intensities: (a) $1\times10^{10}$ $Wcm^{-2}$, (b) $1\times10^{11}$ $Wcm^{-2}$, (c) $5\times10^{11}$ $Wcm^{-2}$, (d) $1\times10^{12}$ $Wcm^{-2}$, and (e) $5\times10^{12}$ $Wcm^{-2}$, shown in grey in the background. (f) – (j) Corresponding high-harmonic spectra J(ω), obtained via Fourier transformation of the induced current densities in (a-e). The fundamental carrier frequency ($\omega_0$) and its harmonics are indicated by vertical dashed lines.

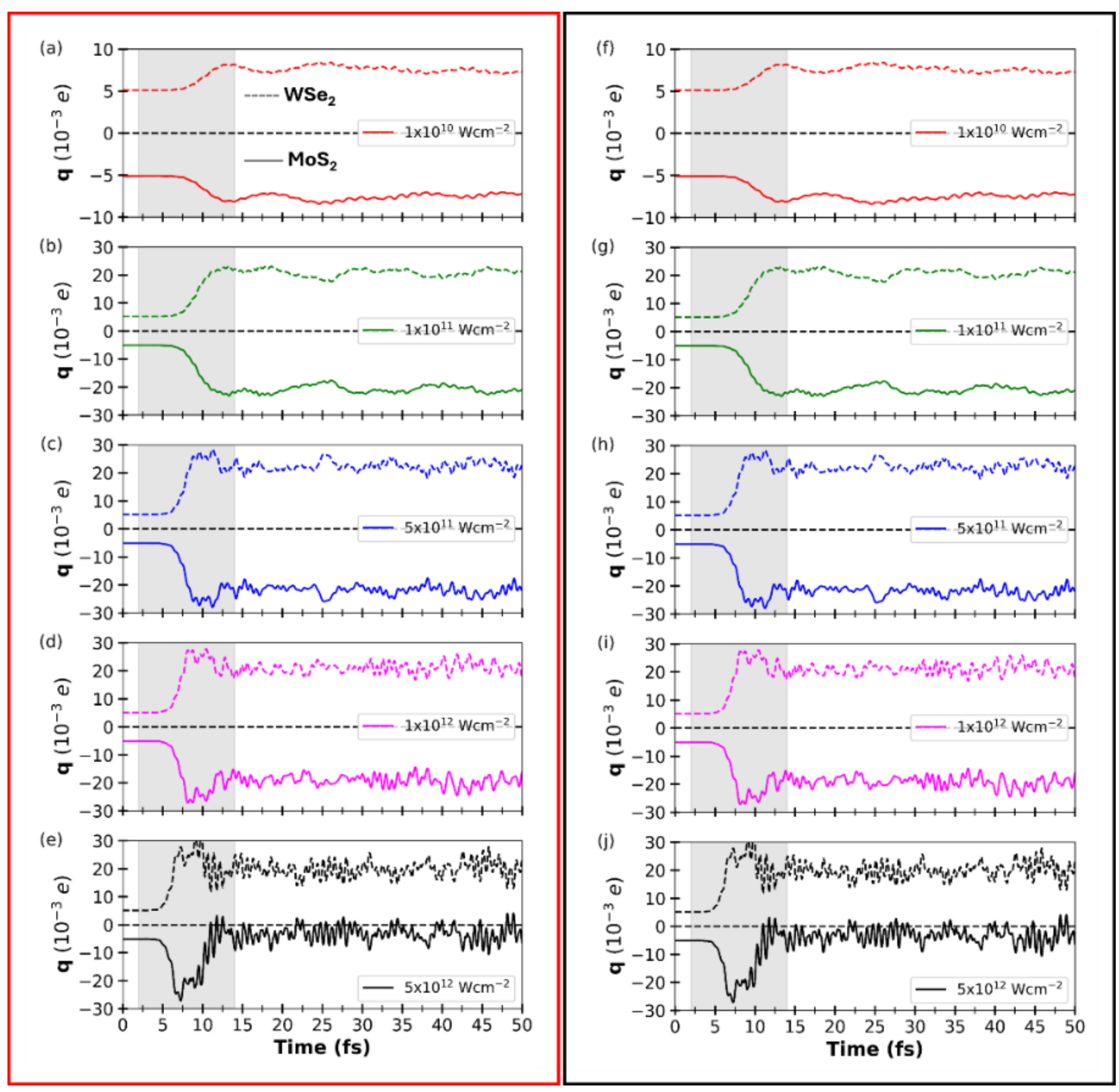

Figure S 6. Time evolution of the layered-resolved atomic charges under the laser intensities of $1\times10^{10}$ $Wcm^{-2}$ (red), $1\times10^{11}$ $Wcm^{-2}$ (green), $5\times10^{11}$ $Wcm^{-2}$ (blue), $1\times10^{12}$ $Wcm^{-2}$ (magenta), and $5\times10^{12}$ $Wcm^{-2}$ (black) for heterobilayer with strain on (a - e) ML $WSe_2$ and (f - j) ML $MoS_2$. The grey area denotes the time window of laser irradiation.

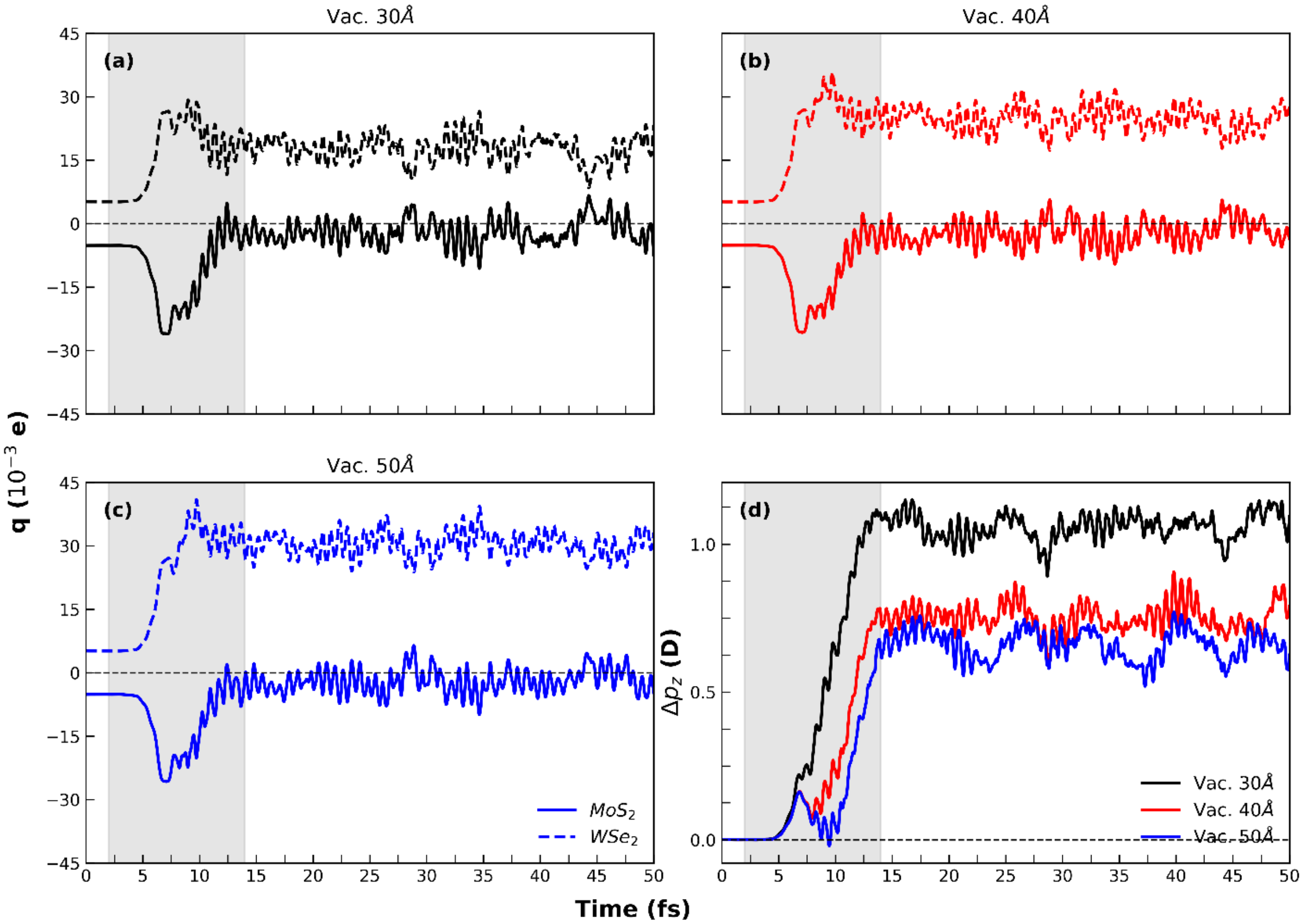

Figure S 7. Time evolution of the layered-resolved atomic charges (a) – (c) and out-of-plane dipole (d) under laser intensity of $5\times10^{12}$ $Wcm^{-2}$ for the heterobilayer with strain distributed to both layers with a minimum vacuum layer of 30 Å (black), 40 Å (red), and 50 Å (blue) thickness. The grey shaded area indicates the time window of laser irradiation.

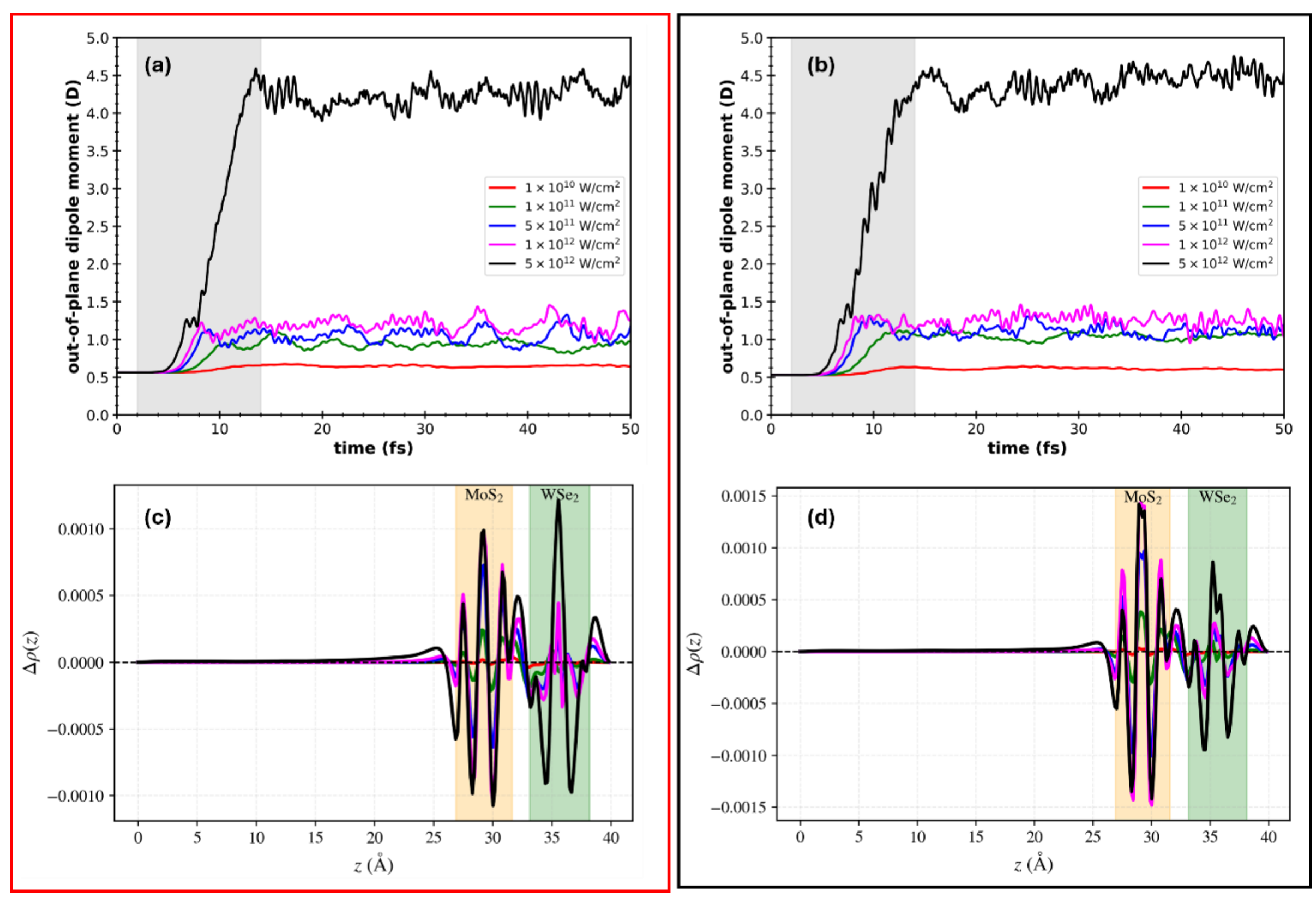

Figure S 8. (a – b) Time evolution of the out-of-plane dipole and (c – d) planar-averaged induced charge density different, Δρ(*z*), under increasing laser intensities: $1\times10^{10}$ Wcm$^{-2}$ (red), $1\times10^{11}$ Wcm$^{-2}$ (green), $5\times10^{11}$ Wcm$^{-2}$ (blue), $1\times10^{12}$ Wcm$^{-2}$ (magenta), and $5\times10^{12}$ Wcm$^{-2}$ (black) for the heterobilayer with strain applied to monolayer $WSe_2$ (left) and monolayer $MoS_2$ (right). The grey shaded area indicates the time window of laser irradiation, while orange and green shaded areas correspond to the regions occupied by $MoS_2$ and $WSe_2$ layers, respectively.